\documentclass[twocolumn,epjc3]{svjour3}
\usepackage{graphics}
\usepackage{multicol}
\usepackage{cuted}
\usepackage{flushend}

\usepackage{microtype}
\usepackage{bbm}
\usepackage{amsmath}
\usepackage{mathtools}
\usepackage{caption}
\usepackage{subcaption}
\usepackage{amssymb}
\usepackage{mathrsfs}       
\usepackage[pdftex]{graphicx}
\usepackage{pgfplots}
\pgfplotsset{width=10cm,compat=1.9}
\usepackage{xargs}       
\usepackage{wasysym}
\usepackage{engrec}
\usepackage{enumerate}
\usepackage{appendix}
\usepackage{empheq}
\usepackage{float}
\usepackage{stmaryrd}
\usepackage[parfill]{parskip}
\usepackage[pdftex,breaklinks]{hyperref}
\usepackage{titletoc}
\usepackage{makecell}
\usepackage{lscape}
\usepackage{algorithm}
\usepackage{algorithmicx}
\usepackage{tensor}
\usepackage{algpseudocode}
\usepackage{blkarray}
\usepackage{multirow}
\usepackage{bigstrut}
\usetikzlibrary{plotmarks}
\usepackage{comment}

\newcommand{\inv}[1]{\frac 1{#1}}
\newcommand{\ket}[1]{|#1\rangle}
\newcommand{\bra}[1]{\langle#1|}
\newcommand{\braket}[2]{\langle#1|#2\rangle}
\newcommand{\ud}{\mathrm{d}}

\newenvironment{customlegend}[1][]{%
    \begingroup
    \csname pgfplots@init@cleared@structures\endcsname
    \pgfplotsset{#1}%
}{%
    \csname pgfplots@createlegend\endcsname
    \endgroup
}%
\def\addlegendimage{\csname pgfplots@addlegendimage\endcsname}

\begin{document}
\title{Multi-reference many-body perturbation theory for nuclei}
\subtitle{I. Novel PGCM-PT formalism}
\author{
M. Frosini\thanksref{ad:saclay,em:mf} \and 
T. Duguet\thanksref{ad:saclay,ad:kul,em:td} \and 
J.-P. Ebran\thanksref{ad:dam,ad:fakedam,em:jpe} \and 
V. Som\`a\thanksref{ad:saclay,em:vs} }                 

\thankstext{em:mf}{\email{mikael.frosini@cea.fr}}
\thankstext{em:td}{\email{thomas.duguet@cea.fr}}
\thankstext{em:jpe}{\email{jean-paul.ebran@cea.fr}}
\thankstext{em:vs}{\email{vittorio.soma@cea.fr}}

\institute{\label{ad:saclay}
IRFU, CEA, Universit\'e Paris-Saclay, 91191 Gif-sur-Yvette, France 
\and
\label{ad:kul}
KU Leuven, Department of Physics and Astronomy, Instituut voor Kern- en Stralingsfysica, 3001 Leuven, Belgium 
\and
\label{ad:dam}
CEA, DAM, DIF, 91297 Arpajon, France
\and
\label{ad:fakedam}
Universit\'e Paris-Saclay, CEA, Laboratoire Mati\`ere en Conditions Extr\^emes, 91680 Bruy\`eres-le-Ch\^atel, France}

\date{Received: \today{} / Revised version: date}

\maketitle
%
%
\begin{abstract}
Perturbative and non-perturbative expansion methods already constitute a tool of choice to perform ab initio calculations over a significant part of the nuclear chart. In this context, the categories of accessible nuclei directly reflect the class of unperturbed state employed in the formulation of the expansion. The present work generalizes to the nuclear many-body context the versatile method of Ref.~\cite{burton20a} by formulating a perturbative expansion on top of a multi-reference unperturbed state mixing deformed non-orthogonal Bogoliubov vacua, i.e. a state obtained from the projected generator coordinate method (PGCM). Particular attention is paid to the part of the mixing taking care of the symmetry restoration, showing that it can be exactly contracted throughout the expansion, thus reducing significantly the dimensionality of the linear problem to be solved to extract perturbative corrections.

While the novel expansion method, coined as PGCM-PT, reduces to the PGCM at lowest order, it reduces to single-reference perturbation theories in appropriate limits. Based on a PGCM unperturbed state  capturing (strong) static correlations in a versatile and efficient fashion, PGCM-PT is indistinctly applicable to doubly closed-shell, singly open-shell and doubly open-shell nuclei. The remaining (weak) dynamical correlations are brought consistently through  perturbative corrections. This symmetry-conserving multi-reference perturbation theory is state-specific and applies to both ground and excited PGCM unperturbed states, thus correcting each state belonging to the low-lying spectrum of the system under study.

The present paper is the first in a series of three and discusses the PGCM-PT formalism in detail. The second paper displays numerical zeroth-order results, i.e. the outcome of PGCM calculations. Second-order, i.e. PGCM-PT(2), calculations performed in both closed- and open-shell nuclei are the object of the third paper.
\end{abstract}

\section{Introduction}
\label{intro}

Given the nuclear Hamiltonian\footnote{The initial nuclear Hamiltonian is typically produced within the frame of chiral effective field theory ($\chi$EFT)~\cite{Epelbaum:2008ga,Epelbaum:2019jbv,Machleidt:2020vzm}. Furthermore, before entering as an input to the presently developed many-body formalism, the Hamiltonian is meant to be evolved via a free-space similarity renormalization group transformation~\cite{Bogner:2009bt}. As an option, and as will be elaborated on in the third paper of the series, one can further pre-process the Hamiltonian via an in-medium similarity renormalization group transformation of single-reference~\cite{Tsukiyama:2010rj,Hergert:2012nb} or multi-reference~\cite{Hergert:2014iaa} types depending on the closed- or open-shell character of the system under study.} $H$, ab initio nuclear structure calculations seek, for as many nuclei as possible, an approximate solution of $A$-body Schr{\"o}dinger's eigenvalue equation
\begin{align}
H | \Psi^{\sigma}_{\mu} \rangle &=  E^{\tilde{\sigma}}_{\mu} \, | \Psi^{\sigma}_{\mu} \rangle \, , \label{schroedevolstate}
\end{align}
that is as accurate as possible. In Eq.~\eqref{schroedevolstate}, $\mu$ denotes a principal quantum number whereas \(\sigma\equiv(\text{J} \text{M} \Pi \text{N} \text{Z})\equiv (\tilde{\sigma}M)\) collects the set of symmetry quantum numbers labelling the many-body states, i.e. the angular momentum J and its projection M, the parity $\Pi$ as well as neutron N and proton Z numbers. The $M$-independence of the eigenenergies $E^{\tilde{\sigma}}_{\mu}$ and the symmetry quantum numbers carried by the eigenstates are a testimony of the symmetry group 
\begin{align}
\text{G}_{H} \equiv \{R(\theta), \theta \in  D_{\text{G}}\}
\end{align}
of the Hamiltonian, i.e.,
\begin{align}
[H,R(\theta)]=0 \, , \, \forall \theta \, ,
\end{align}
which plays a key role in the present context\footnote{The characteristics of $\text{G}_{H}$ and the definitions of the quantities associated with it used throughout the present work are detailed in App.~\ref{symgroup}.}.

\begin{figure}
    \centering
    \includegraphics[width=0.4\textwidth]{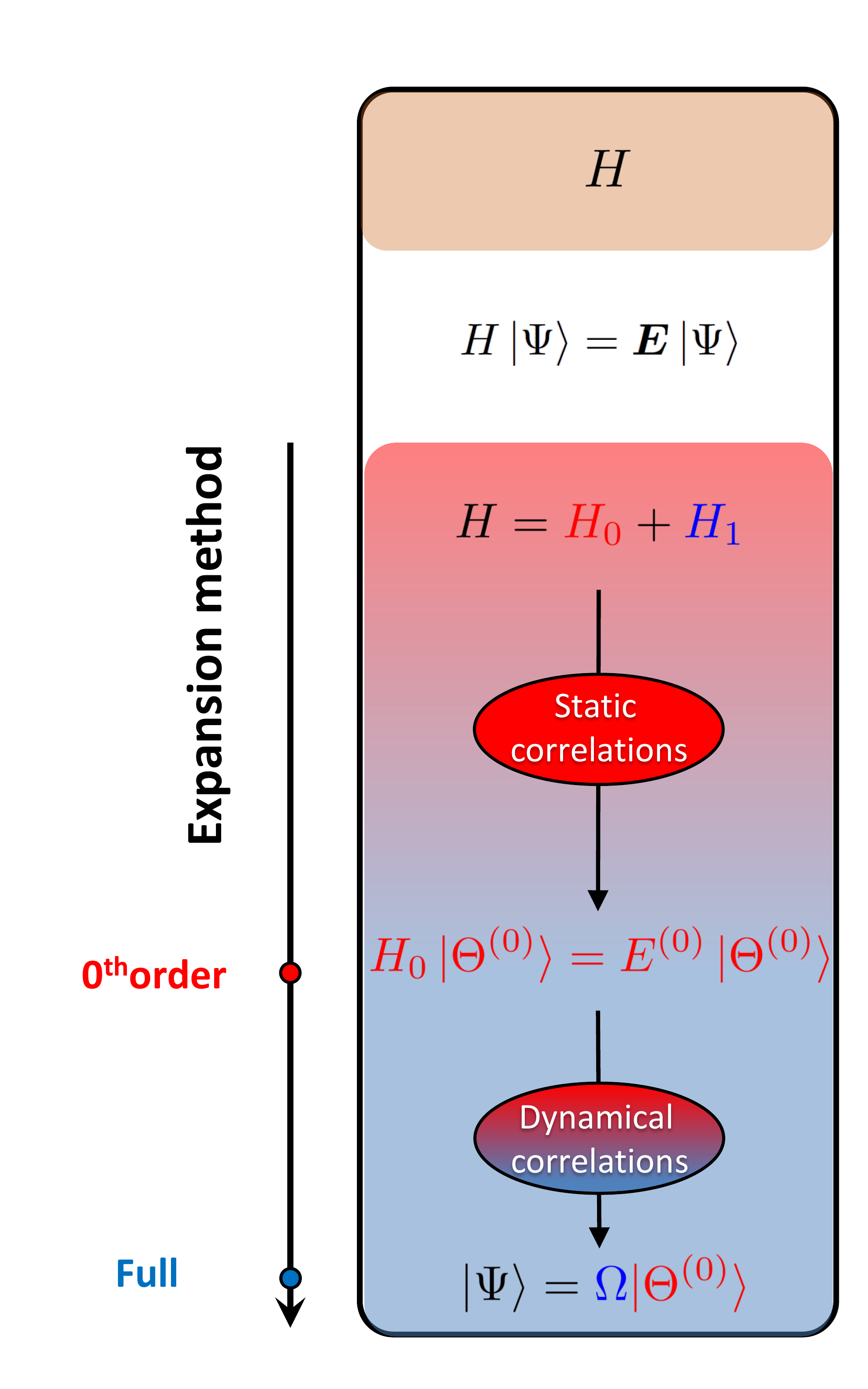}
    \caption{(color online) Schematic illustration of the workflow of expansion many-body methods based on a given input Hamiltonian $H$. While the unperturbed state must be capable of capturing so-called static correlations (if any), the expansion on top of it typically focuses on grasping so-called dynamical correlations (either perturbatively or non-perturbatively).}
    \label{fig:expansion_method}
\end{figure}

The breaking of ab initio calculations away from so-called p-shell nuclei over the last fifteen years has essentially been due to the development and implementation of so-called {\it expansion many-body methods}. Generically, these methods rely on a partitioning of the Hamiltonian
\begin{align}
H &=  H_0 + H_1 \,  \label{partitioninginitial}
\end{align}
chosen such that (at least) one appropriate eigenstate $| \Theta^{\sigma}_{\mu} \rangle$ of $H_0$ is known, i.e. 
\begin{align}
H_0 | \Theta^{\sigma}_{\mu} \rangle &=  E^{\tilde{\sigma}(0)}_{\mu} \, | \Theta^{\sigma}_{\mu} \rangle \, . \label{eignvaluerefstate}
\end{align}
Given this state, the so-called {\it unperturbed state}, expansion methods aim at finding an efficient way to connect it to a target eigenstate $| \Psi^{\sigma}_{\mu} \rangle$ of $H$. This connection is formally achieved via the so-called {\it wave operator}, i.e.
\begin{align}
| \Psi^{\sigma}_{\mu} \rangle &\equiv \Omega_{[\tilde{\sigma},\mu,H_1]} | \Theta^{\sigma}_{\mu} \rangle \, , \label{waveoperator}
\end{align}
which is state specific and carries the complete effect of the residual interaction $H_1$.  This two-step procedure is schematically illustrated in Fig.~\ref{fig:expansion_method}.

Two ingredients characterize a given expansion method
\begin{enumerate}
\item the nature of the partitioning and of the associated unperturbed state,
\item the rationale behind the construction, i.e the expansion and truncation, of the wave operator.
\end{enumerate}
The construction of the wave operator is typically realized via either perturbative~\cite{Tichai:2020dna} or non-perturbative~\cite{Hergert:2020bxy} techniques, i.e. by either expanding $\Omega_{[\tilde{\sigma},\mu,H_1]}$ as a power series in $H_1$ or by organizing the series as a more elaborate function of the residual interaction. Independently of this, the nature and the {\it reach} of the expansion is first and foremost determined by the {\it class} of unperturbed state used, which is itself governed by two main characteristics. The first feature relates to whether $| \Theta^{\sigma}_{\mu} \rangle$ is a pure product state or a linear combination of product states. In the former case, the method is said to be of single reference (SR) nature. In the latter case, the method is said to be of multi-reference (MR) character. One typically needs to transition from the former to the latter whenever the system is (nearly) degenerate and displays strong static correlations making the SR expansion singular, e.g. going from closed-shell to open-shell nuclei. Standard MR unperturbed states are linear combinations of orthonormal Slater determinants spanning a so-called valence/active space. This is indeed the case of the sole MR method implemented to date in nuclear physics, i.e. multi-configuration perturbation theory (MCPT)~\cite{Tichai:2017rqe}. In the present work, the goal is to generalize to the nuclear context the MR perturbation theory developed in Ref.~\cite{burton20a} in which the unperturbed state is built as a linear combination of {\it non-orthogonal} Slater determinants.

In addition to the SR or MR nature of the unperturbed state, a second, key feature relates to the symmetry conserving or non-conserving character of the partitioning in Eq.~\eqref{partitioninginitial}. While expansion methods typically build on a symmetry-conserving scheme~\cite{Shav09MBmethod} as implied by Eq.~\eqref{eignvaluerefstate}, a given SR method can be generalized to a symmetry non-conserving formulation in which 
\begin{subequations}
\label{sympartitioning}
\begin{align}
\left[H_0,R(\theta)\right]&\neq 0 \, , \label{sympartitioning1} \\
\left[H_1,R(\theta)\right]&\neq 0 \, . \label{sympartitioning2}
\end{align}
\end{subequations}
As a result, the unperturbed state is said to be {\it symmetry breaking}, i.e. it loses (some of) the symmetry quantum numbers characterizing the eigenstates of $H$ and rather carries a non-zero order parameter $\varrho\equiv qe^{i\theta}$ whose norm $q$ quantifies the extent of the symmetry breaking. In this case, the unperturbed state is written as $| \Theta_{\mu}(q) \rangle$. Breaking symmetries is employed to capture strong static correlations {\it without} having to resort to MR methods that are typically formally and numerically more involved. While such a philosophy is indeed very powerful~\cite{Soma:2011aj,Soma:2013xha,Signoracci:2014dia,Henderson:2014vka,Tichai:2018vjc,Arthuis:2018yoo,Demol:2020mzd,Soma:2020dyc,Tichai:2021ewr}, it carries the loss of good symmetries over to the (approximation of) $| \Psi^{\tilde\sigma}_{\mu} \rangle$ generated through the expansion method as soon as the wave operator is truncated in Eq.~\eqref{waveoperator}. 

In this context, the natural question is whether broken symmetries can be restored\footnote{While the full wave operator restores broken symmetries, it is always truncated in actual calculations such that the formal restoration obtained in the exact limit is of no practical help.} at any given truncation order whenever a symmetry-breaking SR partitioning is used. Doing so amounts to incorporating large amplitude fluctuations of the {\it phase} $\theta$ of the order parameter characterizing the broken symmetry. While it is indeed straightforward to do so for the unperturbed product state, i.e. at zeroth order in the many-body expansion, by acting with a symmetry projection operator~\cite{RiSc80}, the generalization of symmetry-breaking SR expansion methods to restore the symmetries at any finite truncation order has only attracted serious attention recently, both for perturbation~\cite{Duguet:2014jja,Duguet:2015yle,Arthuis:2020tjz} and coupled cluster~\cite{Duguet:2014jja,Duguet:2015yle,qiu17a} theories\footnote{The a posteriori action of a projector onto a second-order perturbative state was investigated at some point in nuclear physics~\cite{peierls73a,atalay73a,atalay74a,atalay75a,atalay78a} and quantum chemistry~\cite{schlegel86a,schlegel88a,knowles88a} but not pursued since. These methods relied on L\"{o}wdin's representation of the spin projector~\cite{lowdin55c}, often approximating it to only remove the next highest spin.}. These approaches follow a philosophy that can be coined as "partition-and-project-expand" in which the action of the symmetry projector itself needs to be expanded and truncated as soon as one goes beyond zeroth order in the expansion of the wave operator. After showing promising results in the context of solvable/toy models~\cite{qiu17a,Qiu:2018edx}, the novel projected coupled cluster (PCC) formalism has been very recently implemented and tested with success for realistic nuclear structure calculations~\cite{sun21a}. 

The present work wishes to promote an alternative strategy through a novel MR perturbation theory that, while building the unperturbed state out of symmetry-breaking product states, does rather follow a "project-partition-and-expand" philosophy, i.e. relies on a symmetry-conserving partitioning, such that symmetries are exactly maintained all throughout the expansion. Furthermore, the approach is not only capable of incorporating large amplitude fluctuations of the phase of $\varrho$ into the unperturbed state, but also of including large amplitude fluctuations of its norm $q$, which constitutes an efficient asset to fully capture static correlations. 

The present work is made out of three consecutive articles, hereafter coined as Paper I, Paper~II~\cite{paperII} and Paper~III~\cite{paperIII}. Paper I is dedicated to laying out the PGCM-PT formalism whereas Papers II and III present results of the first numerical applications in both closed- and open-shell nuclei. While Paper II focuses on zeroth-order calculations, second-order results are presented and characterized in Paper III.

Paper I is organized as follows. Section~\ref{formal_PT} briefly summarizes formal Rayleigh-Schr{\"o}dinger's perturbation theory~\cite{Shav09MBmethod} underlying the subsequent formulation of the multi-reference perturbation theory of present interest that is formulated in details in Sec.~\ref{PGCMPT_formalism}. Section~\ref{conclusions} 
is dedicated to discussions and conclusions. While the bulk of the paper is restricted to the generic formulation of the many-body method, several appendices provide all algebraic details necessary to the actual implementation of the approach.

\section{Formal perturbation theory}
\label{formal_PT}

\subsection{Set up}
\label{formal_PT_set up}

The present work focuses on the perturbative expansion of the wave operator. Starting from Eqs.~\eqref{partitioninginitial}-\eqref{eignvaluerefstate}, the two projectors in direct sum\footnote{The two hermitian operators fulfill $({\cal P}^{\tilde{\sigma}}_{\mu})^2={\cal P}^{\tilde{\sigma}}_{\mu}$, $({\cal Q}^{\tilde{\sigma}}_{\mu})^2={\cal Q}^{\tilde{\sigma}}_{\mu}$ and ${\cal P}^{\tilde{\sigma}}_{\mu}{\cal Q}^{\tilde{\sigma}}_{\mu}={\cal Q}^{\tilde{\sigma}}_{\mu}{\cal P}^{\tilde{\sigma}}_{\mu}=0$ such that ${\cal P}^{\tilde{\sigma}}_{\mu}+{\cal Q}^{\tilde{\sigma}}_{\mu}=1$.}
\begin{subequations}
\label{projectorsFockspace}
\begin{align}
{\cal P}^{\tilde{\sigma}}_{\mu} &\equiv  \sum_{K} | \Theta^{\tilde{\sigma}K}_{\mu} \rangle \langle \Theta^{\tilde{\sigma}K}_{\mu} | \, , \label{projectorsFockspace1} \\
{\cal Q}^{\tilde{\sigma}}_{\mu} &\equiv  1 - {\cal P}^{\tilde{\sigma}}_{\mu} \, , \label{projectorsFockspace2}
\end{align}
\end{subequations}
are introduced. The operator ${\cal P}^{\tilde{\sigma}}_{\mu}$ projects on the eigen subspace of $H_0$ spanned by the unperturbed state along with the degenerate states obtained via symmetry transformations, i.e. belonging to the same irreducible representation (IRREP) of the symmetry group. This constitutes the so-called ${\cal P}$ space. The operator ${\cal Q}^{\tilde{\sigma}}_{\mu}$ projects onto the complementary orthogonal subspace, the so-called ${\cal Q}$ space. In the present context, the eigenstates spanning the latter eigen subspace of $H_0$ are {\it not} assumed to be known explicitly.

While the ingredients introduced above are state specific and, as such, depend on the quantum numbers $(\mu,\tilde{\sigma})$, those labels are dropped for the time being to lighten the notations. Consequently, the targeted eigenstate and energy of the full Hamiltonian are written as $| \Psi \rangle$ and $E$, respectively, whereas the unperturbed state and energy are denoted as $| \Theta^{(0)} \rangle$ and $E^{(0)}$ to typify that they act as zeroth-order quantities in the perturbative expansion designed below. The projectors are simply denoted as ${\cal P}$ and ${\cal Q}$.

The goal is to compute the perturbative corrections to both $| \Theta^{(0)} \rangle$ and $E^{(0)}$ such that 
\begin{subequations}
\label{perturbative_expansions}
\begin{align}
| \Psi \rangle &\equiv  \sum_{k=0}^{\infty} | \Theta^{(k)} \rangle \, , \label{perturbative_expansions1} \\
E &\equiv \sum_{k=0}^{\infty} E^{(k)} \, , \label{perturbative_expansions2}
\end{align}
\end{subequations}
where the superscript $k$ indicates that the corresponding quantity is proportional to the $k^{\text{th}}$ power of $H_1$. This expansion is defined using the so-called {\it intermediate normalization}, i.e.
\begin{align}
\langle \Theta^{(0)} | \Theta^{(k)} \rangle  = 0 \,\,\, , \,\,\, \forall k \geq 1 \, , \label{orthogonality_PT}
\end{align}
such that
\begin{align}
\langle \Theta^{(0)} | \Psi \rangle  = 1 \, . \label{intermediate_normalization}
\end{align}

\subsection{Perturbative expansion}
\label{formal_PT_set up}

Rayleigh-Schr{\"o}dinger perturbation theory~\cite{Shav09MBmethod} allows one to first expand the exact state and energy as
\begin{subequations}
\label{expansion}
\begin{align}
| \Psi \rangle &\equiv  \sum_{m=0}^{\infty} \left(X^{-1} Y\right)^{m} | \Theta^{(0)} \rangle \, , \label{expansions1} \\
E -E^{(0)}&\equiv \sum_{m=0}^{\infty} \langle \Theta^{(0)} | H_1 \left(X^{-1} Y\right)^{m} | \Theta^{(0)} \rangle \, ,
\label{expansions2}
\end{align}
\end{subequations}
where
\begin{subequations}
\label{defmatrices}
\begin{align}
X &\equiv {\cal Q}\left(H_0-E^{(0)}\right){\cal Q}  \, , \label{defmatrices1} \\
Y &\equiv {\cal Q} \left(E-E^{(0)}- H_1\right) {\cal Q} \, . \label{defmatrices2}
\end{align}
\end{subequations}
The two series do not yet provide the perturbative corrections to the unperturbed quantities because of the presence of $E-E^{(0)}$ on the right-hand side through $Y$. To identify each perturbative contribution, it is necessary to substitute Eq.~\eqref{expansions2} for each $E -E^{(0)}$ in the right-hand-side of Eq.~\eqref{expansion} iteratively and sort out the terms with equal powers of $H_1$. This procedure leads to\footnote{Starting with $| \Theta^{(3)} \rangle$ and $E^{(4)}$, so-called {\it renormalization terms} arise in addition to the {\it principal term}~\cite{Shav09MBmethod}.}\textsuperscript{,}\footnote{The perturbative expansion of the wave operator formally introduced in Eq.~\eqref{waveoperator} is thus obtained as
\begin{align*}
\Omega_{[\tilde{\sigma},\mu,H_1]} &= 1 \nonumber \\
&-(X^{\tilde{\sigma}}_{\mu})^{-1} {\cal Q}^{\tilde{\sigma}}_{\mu} H_1  \nonumber \\
& +(X^{\tilde{\sigma}}_{\mu})^{-1} {\cal Q}^{\tilde{\sigma}}_{\mu} \bar{H}_1 {\cal Q}^{\tilde{\sigma}}_{\mu} (X^{\tilde{\sigma}}_{\mu})^{-1} {\cal Q}^{\tilde{\sigma}}_{\mu} H_1 \nonumber \\
&  + \ldots
\end{align*}
}
\begin{subequations}
\label{wfPT}
\begin{align}
| \Theta^{(1)} \rangle &= -X^{-1} {\cal Q} H_1 | \Theta^{(0)} \rangle \, , \label{wfPT1} \\
| \Theta^{(2)} \rangle &= +X^{-1} {\cal Q} \bar{H}_1 {\cal Q} X^{-1} {\cal Q} H_1 | \Theta^{(0)} \rangle \, , \label{wfPT2} \\
&\vdots \nonumber
\end{align}
\end{subequations}
and\footnote{Some of the projectors ${\cal Q}$ are redundant but are kept to make the systematic structure of the equations more apparent.}
\begin{subequations}
\label{EPT}
\begin{align}
E^{(1)} &= \langle \Theta^{(0)} | H_1 | \Theta^{(0)} \rangle \, , \label{EPT1} \\
E^{(2)} &= \langle \Theta^{(0)} | H_1 {\cal Q} | \Theta^{(1)} \rangle \nonumber \\
&= -\langle \Theta^{(0)} | H_1 {\cal Q} X^{-1} {\cal Q} H_1  | \Theta^{(0)}  \rangle\, , \label{EPT2} \\
E^{(3)} &=  \langle \Theta^{(0)} | H_1 {\cal Q} | \Theta^{(2)} \rangle \nonumber \\
&= +\langle \Theta^{(0)} |  H_1 {\cal Q} X^{-1} {\cal Q}  \bar{H}_1 {\cal Q} X^{-1} {\cal Q} H_1  | \Theta^{(0)}  \rangle \nonumber \\
& = \langle \Theta^{(1)} | {\cal Q}  \bar{H}_1 {\cal Q} | \Theta^{(1)} \rangle \, , \label{EPT3} \\
&\vdots \nonumber
\end{align}
\end{subequations}
where $\bar{H}_1\equiv H_1 - E^{(1)}$. The total energy of the unperturbed state is defined as
\begin{align}
E_{\text{ref}} &\equiv  \langle \Theta^{(0)} | H | \Theta^{(0)} \rangle = E^{(0)} +  E^{(1)} \, .
\end{align}

\subsection{Computable expression}
\label{formal_PT_computation}

Working algebraic expressions of $| \Theta^{(k)} \rangle$ and $E^{(k)}$ are easily obtained in case $X$ is invertible, i.e. if the eigenstates of $H_0$ in ${\cal Q}$ space are known, which is not the case in the present work. Under closer inspection, one actually needs matrix elements of 
\begin{align}
A &\equiv -X^{-1} {\cal Q} \bar{H}_1  \, , 
\end{align}
noting in passing that ${\cal Q} \bar{H}_1 | \Theta^{(0)} \rangle={\cal Q} H_1 | \Theta^{(0)} \rangle$. Since by definition
\begin{align}
{\cal Q}\left(H_0-E^{(0)}\right){\cal Q} A & =  -{\cal Q} \bar{H}_1 \, , 
\end{align}
the matrix ${\bold A}$ of $A$ is the solution of the system of linear equations
\begin{align}
{\bold M} {\bold A} & =  - \bar{{\bold H}}_1 \, , \label{linearsystemmatrix}
\end{align}
where ${\bold M} \equiv {\bold H_0}-E^{(0)}{\bold 1}$ and where the left matrix index necessarily belongs to ${\cal Q}$ space whereas the right index is either in ${\cal Q}$ or ${\cal P}$ space. In expanded form, the linear system reads, with $i\neq 0$,
\begin{align}
\sum_{k\neq 0} M_{ik} A_{kj} & =  -\left(\bar{H}_1\right)_{ij} \, , \label{linearsystemmatrix2}
\end{align}
where the sum is restricted to ${\cal Q}$-space states. In case one is only interested in $X^{-1}{\cal Q} \bar{H}_1 | \Theta^{(0)} \rangle$, a simpler linear system involving the vectors ${\bold a}$ and ${\bold h_1}$ made out of the first column $A_{k0}$ and $(H_1)_{k0}$ of ${\bold A}$ and ${\bold H_1}$, respectively, needs to be solved, i.e. 
\begin{align}
{\bold M} {\bold a} & =  -{\bold h_1} \, . \label{linearsystemvector}
\end{align}
As discussed in Paper III, a sparse matrix representation of $M$ makes the iterative solution of the linear equation system accessible under certain hypothesis for realistic ab initio nuclear structure calculations.

Given ${\bold A}$, the energy corrections can eventually be computed as
\begin{subequations}
\label{EPTbis}
\begin{align}
E^{(2)} &= \langle \Theta^{(0)} | H_1 A | \Theta^{(0)} \rangle = {\bold h_1^{\dagger}} {\bold a} \, , \label{EPTbis2} \\
E^{(3)} &=  \langle \Theta^{(0)} | H_1 A^2 | \Theta^{(0)} \rangle = {\bold h_1^{\dagger}} {\bold A}{\bold a} = {\bold a^{\dagger}} \bar{{\bold H}}_1 {\bold a} \, , \label{EPTbis3} \\
&\vdots \, \nonumber
\end{align}
\end{subequations}
knowing that $E^{(1)}=\left(H_1\right)_{00}$.

\subsection{Hylleraas functional}
\label{formal_PT_hylleraas}

Formal perturbation theory can be alternatively derived through a variational method due to Hylleraas~\cite{Hylleraas,Shav09MBmethod}. Let us consider a variational ansatz
\begin{align}
| \Xi \rangle &\equiv  | \Theta^{(0)} \rangle + \sum_{k=1}^{\infty} | \Xi^{(k)} \rangle  \, , \label{hylleraas1}
\end{align}
where $\langle \Theta^{(0)} | \Xi^{(k)} \rangle  = 0 \, \forall k \geq 1$ and where the variational component $| \Xi^{(k)} \rangle$ is proportional to $H_1^k$. Computing the expectation of $H$ in $| \Xi \rangle$ and sorting the various orders in $H_1$, Ritz' variational principle leads to
\begin{align}
E \leq&  E^{(0)} + E^{(1)} \nonumber \\
&+ \left[\langle \Xi^{(1)} | {\cal Q} H_1 | \Theta^{(0)} \rangle + \langle \Theta^{(0)}  | H_1 {\cal Q} | \Xi^{(1)}\rangle \right. \nonumber \\
&\,\,\,\,\,\,+ \left. \langle \Xi^{(1)} | {\cal Q} (H_0 - E^{(0)}) {\cal Q} | \Xi^{(1)} \rangle \right] + O(H_1^3)  \, , \label{hylleraas2}
\end{align}
For $E$ to be a minimum of the right-hand side expression for an arbitrary $H_1$, each term associated with a given power of $H_1$ must be either minimal or constant. The sum of the corresponding terms delivers the individual perturbative components $E^{(k)}$ in Eq.~\eqref{perturbative_expansions2} given the uniqueness of the series in powers of $H_1$.

Noting that $E^{(0)}$ and $E^{(1)}$ are free from any variational components, the variational approach starts with the second-order energy correction $E^{(2)}$ that is the minimum of the so-called Hylleraas functional
\begin{align}
L[\Xi^{(1)}] &\equiv  \langle \Xi^{(1)} | {\cal Q} H_1 | \Theta^{(0)} \rangle \nonumber \\
&+ \langle \Theta^{(0)}  | H_1 {\cal Q} | \Xi^{(1)}\rangle \nonumber \\
&+ \langle \Xi^{(1)} | {\cal Q} (H_0 - E^{(0)}) {\cal Q} | \Xi^{(1)} \rangle  \, . \label{hylleraas3}
\end{align}
It is straightforward to realize that the saddle-point of Eq.~\eqref{hylleraas3} is obtained for $| \Xi^{(1)} \rangle = | \Theta^{(1)} \rangle$ solution of Eq.~\eqref{wfPT1}. 

This alternative derivation is of interest because it underlines the fact that the use of an approximate ansatz to the exact solution of Eq.~\eqref{wfPT1} delivers a variational estimate\footnote{One however obtains a variational upper bound of the exact eigen energy if and only if $E^{(0)}$ is the lowest eigenvalue of $H_0$~\cite{Shav09MBmethod}.} of $E^{(2)}$.

\section{PGCM-PT formalism}
\label{PGCMPT_formalism}

The above formal perturbation theory is now specified to the case where the unperturbed state is generated through the projected generator coordinate method (PGCM). The unperturbed state is thus of MR character given that a PGCM state is nothing but a linear combination of non-orthogonal product states whose coefficients result from solving Hill-Wheeler-Griffin's (HWG) secular problem~\cite{RiSc80}, i.e. a generalized many-body eigenvalue problem. The PGCM perturbation theory (PGCM-PT) of present interest adapts to the nuclear many-body problem the MR perturbation theory recently formulated in the context of quantum chemistry~\cite{burton20a} where the reference state arises from a non-orthogonal configuration interaction (NOCI) calculation involving Slater determinants. In order to do so, the method is presently generalized to the mixing of Bogoliubov vacua.

In the present context, PGCM must thus be viewed as the unperturbed, i.e. zeroth-order, limit of the PGCM-PT formalism that is universally applicable, i.e. independently of the closed or open-shell nature of the system and of the ground or excited character of the PGCM state generated though the initial HWG problem. Because PGCM states efficiently capture strong static correlations associated with the spontaneous breaking of symmetries and their restoration as well as with large amplitude collective fluctuations, one is only left with incorporating the remaining weak dynamical correlations, which PGCM-PT offers to do consistently. Because of the incorporation of static correlations into the zeroth-order state, the hope is that nuclear observables associated with a large set of nuclei and quantum states can be sufficiently converged at low orders in PGCM-PT.

\subsection{Hamiltonian}
\label{H}

The Hamiltonian is presently taken to contain two-nucleon interactions only for simplicity
\begin{align}
    H \equiv& T + V   \notag\\
    \equiv& \frac{1}{(1!)^2} \sum_{\substack{a_1\\b_1}} t^{a_1}_{b_1} \, C^{a_1}_{b_1} +\frac{1}{(2!)^2} \sum_{\substack{a_1a_2\\b_1b_2}} v^{a_1a_2}_{b_1b_2} \, C^{a_1a_2}_{b_1b_2} \ ,  \label{originalH}
\end{align}
where $\{c^{\dagger}_a, c_a\}$ denotes an arbitrary one-body Hilbert space ${\cal H}_1$ and where
\begin{equation} 
C^{a_1\cdots a_n}_{b_1\cdots b_n}\equiv c^\dag_{a_1}\cdots c^\dag_{a_n}c_{b_n}\cdots c_{b_1} \label{stringpart}
\end{equation}
defines a string of $n$ particle creation and $n$ particle annihilation operators\footnote{More details regarding the second-quantized form of operators can be found in App.~\ref{operators}.}.

In practice, such a Hamiltonian typically results from a first step in which the three-nucleon operator has been approximated via a rank-reduction method; see Ref.~\cite{Frosini:2021tuj} for details. Still, the many-body formalism presently introduced can be formulated in the presence of explicit three-nucleon interactions without any formal difficulty.

\subsection{PGCM unperturbed state}
\label{partitioning}

\subsubsection{Ansatz}
\label{ansatzPGCM}

A MR PGCM state can be written as
\begin{align}
		\ket{\mathrm \Theta^{\sigma}_{\mu} }
		&\equiv \int\ud q f^{\tilde{\sigma}}_{\mu}(q) P^{\tilde{\sigma}}_{M0} | \Phi(q) \rangle \nonumber \\
		&= \frac{d_{\tilde{\sigma}}}{v_{\text{G}}}  \sum_q f^{\tilde{\sigma}}_{\mu}(q) \sum_\theta
	D_{\text{M0}}^{\tilde{\sigma}\ast}(\theta) | \Phi (q;\theta) \rangle \, , \label{PGCMstate}
\end{align}
where integrals over the collective coordinate $q$ and the rotation angle $\theta$ have been discretized as actually done in a practical calculation. 

In Eq.~\eqref{PGCMstate}, $\text{B}_{q} \equiv \{ \ket{\Phi(q)}; q \in \text{set} \}$ denotes a set of non-orthogonal Bogoliubov states differing by the value of the collective deformation parameter $q$. Such an ansatz is characterized by its capacity to efficiently capture static correlations from a low-dimensional, i.e. containing from several tens to a few hundreds states, configuration mixing at the price of dealing with non-orthogonal vectors. This constitutes a very advantageous feature, especially as the mass A of the system, and thus the dimensionality of the Hilbert space ${\cal H}_{\text{A}}$, grows.

The product states belonging to $\text{B}_{q}$ are typically obtained in a first step by solving repeatedly Hartree-Fock-Boboliubov (HFB) mean-field equations with a Lagrange term associated with a constraining operator\footnote{The generic operator $Q$ can embody several constraining operators such that the collective coordinate $q$ may in fact be multi dimensional.} $Q$ such that the solution satisfies
\begin{align}
\langle \Phi(q) | Q | \Phi(q) \rangle &= q \, . \label{constraint}
\end{align}
The constrained HFB total energy ${\bold H}^{00}(q)$ (see Eq.~\eqref{HFBenergy}) delivers as a function of $q$, the so-called HFB total energy curve (TEC). Details about Bogoliubov states and the associated algebra, as well as  constrained HFB equations, can be found in App. \ref{bogoalgebra}. 
The constraining operator $Q$ is typically defined such that the product states belonging to $\text{B}_q$ break a symmetry of the Hamiltonian as soon as $q\neq 0$. Because physical states must carry good symmetry quantum numbers one acts on $| \Phi(q) \rangle$ with the operator\footnote{The present work is effectively concerned with HFB states that are invariant under spatial rotation around a given symmetry axis. Extending the formulation to the case where $| \Phi(q) \rangle$ does not display such a symmetry poses no formal difficulty but requires a more general projection operator $P^{\sigma}$; see App.~\ref{symgroup} for details.} 
\begin{align}
P^{\tilde{\sigma}}_{M0} &=\, \frac{d_{\tilde{\sigma}}}{v_{\text{G}}} \int_{D_{\text{G}}} \text{d}\theta  D_{\text{M0}}^{\tilde{\sigma}\ast}(\theta) R(\theta)   \label{projectoraxial}
\end{align}
in Eq.~\eqref{PGCMstate} to project the HFB state onto eigenstates of the symmetry operators with eigenvalues \((\tilde{\sigma},M)\). The operator $P^{\tilde{\sigma}}_{M0}$ is expressed in terms of the symmetry rotation operator $R(\theta)$ and the IRREP $D_{\text{MK}}^{\tilde{\sigma}}(\theta)$ of the symmetry group $\text{G}_{H}$. See App.~\ref{symgroup} for a discussion of the actual symmetry group, symmetry quantum numbers and symmetry projector of present interest. 
\begin{figure*}
    \centering
    \includegraphics[width=0.8\textwidth]{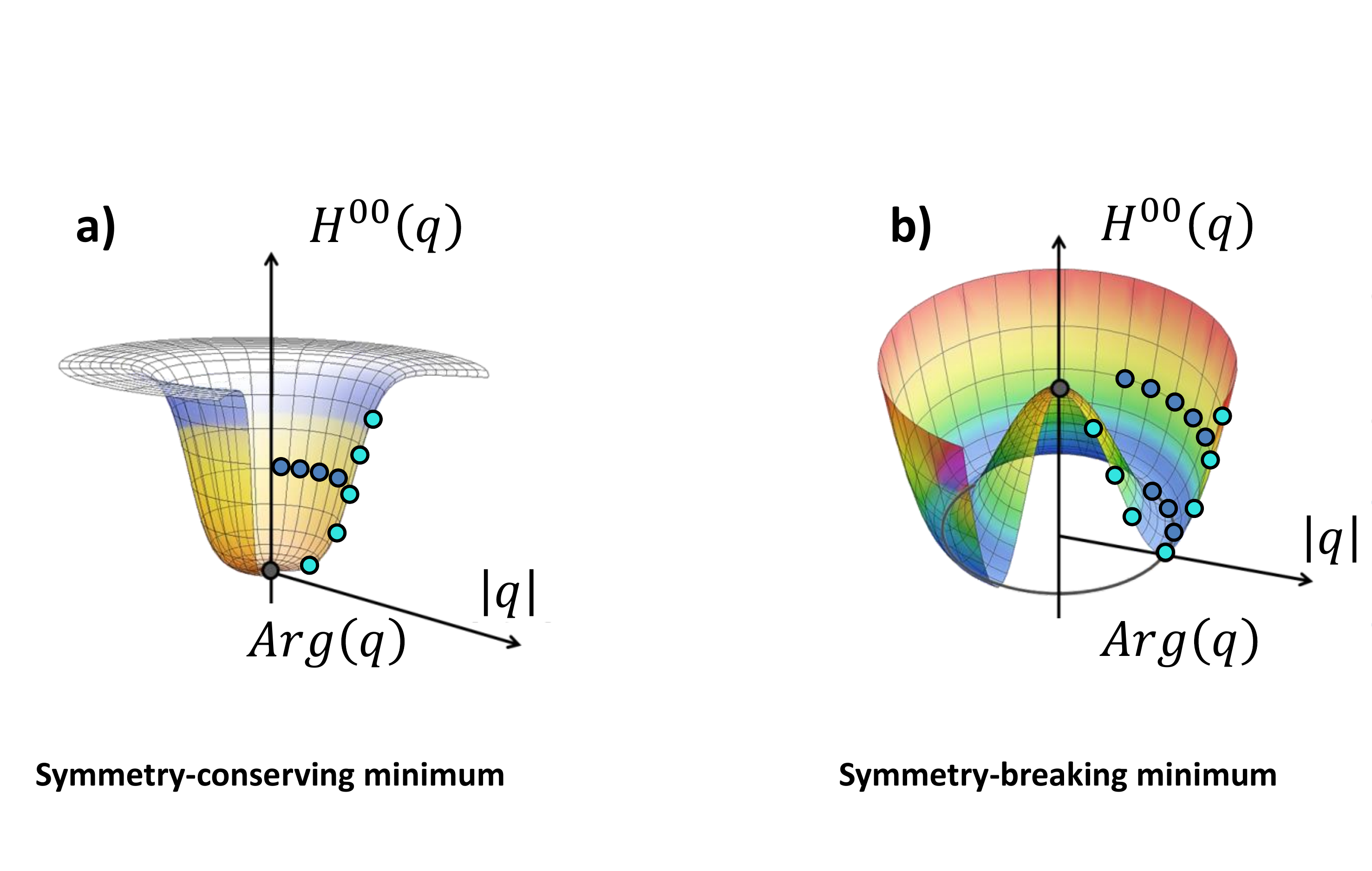}
    \caption{(color online) Schematic representation of the HFB TES ${\bold H}^{00}(q; \theta)$ in the two-dimensional plane associated with the order parameter $\varrho = qe^{i\theta}$ of the (intermediately broken) symmetry. Light (dark) blue circles represent configurations along the $q$ ($\theta$) direction. Left (right) panel: system characterized by a symmetry-conserving (-breaking) HFB minimum.}
    \label{fig:mex_hat_ref}
\end{figure*}

Due to the symmetry projection, the PGCM state is eventually constructed from an extended set $\text{B}_{q; \theta} \equiv \{ \ket{\Phi(q; \theta)}; q \in \text{set} \, \, \text{and} \, \, \theta \in D_{\text{G}} \}$\footnote{Seeing the PGCM state as a configuration mixing of states belonging to $\text{B}_{q; \theta}$ rather than as resulting from the projection of the states belonging $\text{B}_{q}$ allows one to define the SR limit of PGCM-PT via the truncation of the double sum in Eq.~\eqref{PGCMstate} to a single term such that the PGCM unperturbed state reduces to one symmetry-breaking state $| \Phi(q;0) \rangle$.} of Bogoliubov states obtained from $\text{B}_{q}$ by further acting with the symmetry rotation operator, i.e.
\begin{align}
| \Phi(q;\theta) \rangle &\equiv R(\theta) | \Phi(q) \rangle \, . \label{rotated_bogo}
\end{align}
Because $R(\theta) \in \text{G}_{H}$, the HFB total energy surface (TES) ${\bold H}^{00}(q; \theta)$ (Eq.~\eqref{rotatedHFBenergy}) in the two-dimensional plane associated with the order parameter $\varrho = qe^{i\theta}$ of the (intermediately broken) symmetry is independent of the rotation angle $\theta$. This feature is illustrated in Fig.~\ref{fig:mex_hat_ref}.

Mixing states belonging to $\text{B}_{q; \theta}$, PGCM states account for (potentially) large amplitude fluctuations of both the norm $q$ and the angle $\theta$ of the order parameter $\varrho$. Doing so constitutes an efficient way to incorporate strong static correlations and extract at the same time collective excitations associated with both of these fluctuations, i.e. vibrational and rotational excitations along with their coupling. In this context, the coefficients $D_{\text{M0}}^{\tilde{\sigma}}(\theta)$ associated with the symmetry restoration are entirely determined by the group structure\footnote{This is true because the present work is only concerned with HFB states that are invariant under spatial rotation around a given symmetry axis. If not, the configuration mixing with respect to the rotation (i.e. Euler) angles is not entirely fixed by the structure of the group; see App.~\ref{symgroup} for details.}. Consequently, the sole unknowns to be determined are the coefficients $f^{\tilde{\sigma}}_{\mu}(q)$ associated with the mixing over $q$ in Eq.~\ref{PGCMstate}.

\subsubsection{Hill-Wheeler-Griffin's equation}
\label{NOCIeq}

The unknown coefficients $\{f^{\tilde{\sigma}}_{\mu}(q); q \in \text{set} \}$ are determined via the application of Ritz' variational principle
\begin{align}
\frac{\delta}{\delta f^{\tilde{\sigma}\ast}_{\mu}(q)} \frac{\langle \Theta^{\sigma}_{\mu} | H | \Theta^{\sigma}_{\mu} \rangle}{\langle \Theta^{\sigma}_{\mu} | \Theta^{\sigma}_{\mu} \rangle} = 0\, , \label{ritz_PGCM}
\end{align}
that eventually leads to solving HWG's equation\footnote{The diagonalization is performed separately for each value of $\tilde{\sigma}$, i.e. within each IRREP of $\text{G}_{H}$.}
\begin{align}
\sum_{q} H^{\tilde{\sigma}}_{p0q0} \, f^{\tilde{\sigma}}_{\mu}(q) &=  {\cal E}^{\tilde{\sigma}}_{\mu} \sum_{q}  N^{\tilde{\sigma}}_{p0q0} \, f^{\tilde{\sigma}}_{\mu}(q) \, . \label{HWG_equation}
\end{align}
Equation~\ref{HWG_equation} is nothing but a NOCI eigenvalue problem expressed in the set $\text{PB}_{q\tilde{\sigma}} \equiv \{ P^{\tilde{\sigma}}_{00}\ket{\Phi(q)}; q \in \text{set} \}$ of non-orthogonal {\it projected} HFB states. The coefficients $f^{\tilde{\sigma}}_{\mu}(q)$ are presently defined such that the set of PGCM states $\text{PGCM}_{\sigma} \equiv\{\ket{\mathrm \Theta^{\sigma}_{\mu} } ; \mu = 1, 2,\ldots\}$ emerging from Eq.~\eqref{HWG_equation} are (ortho-)normalized. 

Equation~\eqref{HWG_equation} involves so-called {\it operator kernels}\footnote{The two 0 indices in $O^{\tilde{\sigma}}_{p0q0}$ relate to the fact that the ket and the bra denote HFB vacua belonging to  $\text{B}_q$. This notation is necessary to make those kernels consistent with the more general ones $O^{\tilde{\sigma}}_{pIqJ}$ introduced later on in Sec.~\ref{second_order}, which also involve Bogoliubov states obtained via elementary, i.e. quasi-particle, excitations of those belonging to $\text{B}_q$.}
\begin{align}
O^{\tilde{\sigma}}_{p0q0} &\equiv  \bra{\Phi(p)} O P^{\tilde{\sigma}}_{00} \ket{\Phi(q)} \, , \label{kernels_PGCM}
\end{align}
whose explicit algebraic expressions in terms of input quantities are worked out in App.~\ref{PGCMPT2matel}. The kernel associated with the identity operator $N\equiv 1$ is the norm kernel denoted as $N^{\tilde{\sigma}}_{p0q0}$.

The more the vectors in $\text{PB}_{q\tilde{\sigma}}$ are linearly dependent, the more the generalized eigenvalue problem of Eq.~\eqref{HWG_equation} tends to be singular. In order to avoid numerical instabilities, singular eigenvalues must be removed. The standard method to deal with the problem, which is feasible for the manageable number of states in $\text{PB}_{q\tilde{\sigma}}$, is recalled in Paper II.

\subsection{Partitioning}
\label{sec_partitioning}

Now that the nature of the MR unperturbed state $| \Theta^{\sigma}_{\mu} \rangle$ has been detailed, the formal perturbation theory developed in Sec.~\ref{formal_PT} can be explicitly applied. The formulation starts with the choice of an appropriate partitioning of the Hamiltonian (Eq.~\ref{partitioninginitial}), i.e. by defining an unperturbed Hamiltonian $H_0$ of which $| \Theta^{\sigma}_{\mu} \rangle$ is an eigenstate. 

\subsubsection{Definition}
\label{def_part}

The goal is to design $H_0$ such that PGCM-PT reduces to standard M{\o}ller-Plesset MBPT whenever the PGCM unperturbed state reduces to a single (unconstrained) Hartree-Fock (HF) Slater determinant\footnote{This limit is discussed in Sec.~\ref{SR_sym_lim_part}. The more subtle cases where the PGCM unperturbed state reduces to a single {\it constrained} HF Slater determinant or a single Bogoliubov state are also discussed.}. To achieve this goal, one introduces the state-specific partitioning
\begin{align}
H_0 \equiv {\cal P}^{\tilde{\sigma}}_{\mu} F_{[| \Theta \rangle]} {\cal P}^{\tilde{\sigma}}_{\mu} + {\cal Q}^{\tilde{\sigma}}_{\mu} F_{[| \Theta \rangle]} {\cal Q}^{\tilde{\sigma}}_{\mu} \, , \label{def_H0}
\end{align}
where the one-body operator
\begin{subequations}
\label{def_F}
\begin{align}
F_{[| \Theta \rangle]} &\equiv  \sum_{a_1b_1} f^{a_1}_{b_1}[| \Theta \rangle] \, C^{a_1}_{b_1} \, , \label{def_F1} \\
f^{a_1}_{b_1}[| \Theta \rangle] &\equiv t^{a_1}_{b_1} + \sum_{a_2 b_2} v^{a_1a_2}_{b_1b_2} \, 	\left[{\rho^{\Theta}}\right]^{b_2}_{a_2} \, , \label{def_F2}
\end{align}
\end{subequations}
involves the convolution of the two-body interaction with a {\it symmetry-invariant} one-body density matrix\footnote{In case $| \Theta \rangle$ were to denote the exact ground-state of the system, $F_{[| \Theta \rangle]}$ would be nothing else but the so-called Baranger one-body Hamiltonian~\cite{baranger70a}, which is the energy-independent part of the one-nucleon self-energy in self-consistent Green's function theory~\cite{Duguet:2014tua}.}
\begin{equation}
	\left[{\rho^{\Theta}}\right]^{b_1}_{a_1} \equiv \frac{\langle \Theta | C^{a_1}_{b_1} | \Theta \rangle}{\langle \Theta | \Theta \rangle} \, , \label{deflbodydensmat}
\end{equation}
i.e. a one-body density matrix computed from a symmetry-conserving state $| \Theta \rangle$\footnote{In the present work, a symmetry-conserving state represents a state whose associated one-body density matrix is symmetry-invariant, i.e. belongs to the {\it trivial} IRREP of $G_H$. While for the SU(2) group this requires the many-body state itself to be symmetry invariant, i.e. to be a $J=0$ state, for the U(1) group this condition is automatically satisfied for the {\it normal} one-body density matrix.}. 

As soon as the PGCM unperturbed state $| \Theta^{\sigma}_{\mu} \rangle$ is not symmetry-conserving, e.g., it corresponds to an excited state  $| \Theta^{\sigma}_{\mu} \rangle$ of an even-even nucleus with $J\neq 0$, the one-body operator $F_{[| \Theta \rangle]}$ must necessarily be built from a different state $| \Theta \rangle$. In this situation, it is natural to employ the corresponding symmetry-conserving ground state\footnote{For odd-even or odd-odd nuclei eigenstates, the symmetry-invariant density matrix associated with a {\it fake} odd system described in terms of, e.g., a statistical mixture~\cite{Duguet01a,PerezMartin08a} can typically be envisioned.}. Contrarily, whenever the PGCM unperturbed state $| \Theta^{\sigma}_{\mu} \rangle$ is symmetry-conserving, e.g. for the ground state of an even-even system, it is natural to choose it\footnote{The explicit expression of the one-body density matrix of a PGCM state can be found in App. B of Ref.~\cite{Frosini:2021tuj}.}, i.e. to take $| \Theta \rangle\equiv | \Theta^{\sigma}_{\mu} \rangle$, to build $F_{[| \Theta \rangle]}$.

\subsubsection{Eigenstates of $H_0$}
\label{eigen_H0}

Introducing\footnote{The dependence of $E^{\tilde{\sigma}(0)}_{\mu}$ on $| \Theta \rangle$ is dropped for simplicity.} the $M$-independent unperturbed energy
\begin{align}
E^{\tilde{\sigma}(0)}_{\mu} &\equiv \langle \Theta^{\tilde{\sigma}M}_{\mu} | F_{[| \Theta \rangle]} | \Theta^{\tilde{\sigma}M}_{\mu} \rangle =\sum_{a_1 b_1} f^{a_1}_{b_1} \left[\ket\Theta\right]\, \left[{\rho^{\Theta^{\tilde{\sigma}0}_{\mu}}}\right]^{b_1}_{a_1}  \, , \label{value_E0}
\end{align}
one can write
\begin{align}
H_0 \equiv E^{\tilde{\sigma}(0)}_{\mu} {\cal P}^{\tilde{\sigma}}_{\mu}  + {\cal Q}^{\tilde{\sigma}}_{\mu} F_{[| \Theta \rangle]} {\cal Q}^{\tilde{\sigma}}_{\mu} \, , \label{def_H02}
\end{align}
such that the PGCM state is by construction, and independently of the detailed definition of $| \Theta \rangle$, an eigenstate of $H_0$ with eigenvalue $E^{\tilde{\sigma}(0)}_{\mu}$.

Because the PGCM unpertubed state is a linear combination of non-orthogonal product states, there is no preferred one-body basis of ${\cal H}_1$ that can be used to represent this state or the other eigenstates of $H_0$ conveniently. This feature reflects the fact that, while $F_{[\ket\Theta]}$ is a one-body operator with an explicit second-quantized representation, $H_0$ is a genuine {\it many-body} operator with no simple second-quantized form. This further results into the fact that the other eigenstates of $H_0$ are not accessible via excitations of the unperturbed state that are simply built from a given set of one-body creation and annihilation operators. As a matter of fact, $| \Theta^{\sigma}_{\mu} \rangle$ constitutes the only eigenstate of $H_0$ at hand given that no explicit eigen representation of $H_0$ in the complementary ${\cal Q}^{\tilde{\sigma}}_{\mu}$ space is trivially accessible. This difficulty was anticipated in Sec.~\ref{formal_PT} where formal perturbation theory was presented {\it without} assuming that such an eigen-representation was available. The practical consequences for the second-order implementation, i.e. PGCM-PT(2), are discussed in detail below in Sec.~\ref{1storderWF}.

\subsubsection{Symmetries}
\label{sym_part}

Being built from a symmetry-invariant one-body density matrix, $F_{[| \Theta \rangle]}$ is symmetry invariant, i.e. it belongs to the trivial IRREP of $G_H$ such that
\begin{align}
[F,R(\theta)] &= 0 \,  ,\, \, \forall \, \theta \, , \label{F_sym}
\end{align}
which is notably responsible for the $M$-independence of $E^{\tilde{\sigma}(0)}_{\mu}$ in Eq.~\eqref{value_E0}. Using Eqs.~\eqref{unitarity} and~\eqref{eq:ten:def}, one can further prove that 
\begin{align}
[{\cal P}^{\tilde{\sigma}}_{\mu},R(\theta)] &= 0 \, , \, \, \forall \, \theta \, , \label{part_comm}
\end{align}
and thus similarly for ${\cal Q}^{\tilde{\sigma}}_{\mu}$. Consequently, $H_0$ itself, and thus $H_1$, are scalars with respect to $G_H$, i.e. 
\begin{align}
[H_0,R(\theta)] &= [H_1,R(\theta)]  = 0 \, \, \, , \forall \, \theta \, , \label{H0_comm}
\end{align}
such that the partitioning is indeed symmetry conserving. Consequently, the eigenstates of $H_0$, most of which are not known explicitly as discussed in the previous section, carry the symmetry quantum numbers $\sigma = (\tilde{\sigma},M)$.

The unperturbed PGCM state $\ket{\mathrm \Theta^{\tilde{\sigma}M}_{\mu} }$ introduced in Eq.~\eqref{PGCMstate} can be rewritten as 
\begin{align}
| \Theta^{(0)} \rangle &\equiv  P^{\tilde{\sigma}}_{M0} | \bar{\Theta}^{(0)} \rangle = P^{\tilde{\sigma}}_{M0} (P^{\tilde{\sigma}}_{00} | \bar{\Theta}^{(0)} \rangle )\, , \label{PGCM_new}
\end{align}
where $| \bar{\Theta}^{(0)} \rangle$ is an eigenstate of $J_z$ with eigenvalue $M=0$. Exploiting the scalar character of $H_0$, $H_1$ and ${\cal Q}^{\tilde{\sigma}}_{\mu}$, the $k^{\text{th}}$-order perturbed state (Eq.~\eqref{wfPT}) can similarly be shown to read
\begin{align}
| \Theta^{(k)} \rangle & \equiv  P^{\tilde{\sigma}}_{M0} | \bar{\Theta}^{(k)} \rangle   = P^{\tilde{\sigma}}_{M0} ( P^{\tilde{\sigma}}_{00}| \bar{\Theta}^{(k)} \rangle) \, . \label{kth_order_new}
\end{align}
Thus, one can choose to solve for $| \Theta^{(0)} \rangle = \ket{\mathrm \Theta^{\tilde{\sigma}0}_{\mu} }$ and thus for $| \Theta^{(k)} \rangle = P^{\tilde{\sigma}}_{00}| \bar{\Theta}^{(k)} \rangle$. Further acting a posteriori on the obtained solution with the operator $P^{\tilde{\sigma}}_{M0}$ generates all the associated states of the IRREP.

Furthermore, the intermediate states entering $| \bar{\Theta}^{(k)} \rangle$ as a result of the repeated action of $H_1$ and $X^{-1}$ (see Eq.~\eqref{wfPT}) on $| \bar{\Theta}^{(0)} \rangle$ also carry $M=0$; i.e. PGCM-PT can effectively be implemented without any loss of generality within the restricted $\sigma = (\tilde{\sigma},M=0)$ sub-block, i.e. without ever connecting to states with $M\neq 0$.

\subsubsection{$U(1)$-conserving single-reference limit}
\label{SR_sym_lim_part}

The presently developed PGCM-PT can be investigated in the limit where the unperturbed state becomes of single-reference nature. It corresponds to reducing the set $B_{q \theta}$ in Eq.~\eqref{PGCMstate} to a single HF(B) state such that PGCM-PT must exhibit some connection with single-reference (B)MBPT~\cite{Tichai:2020dna} in this limit. In fact, the characteristics of the single-reference limit depend on the symmetry properties of the unperturbed product state, which requires to distinguish two cases.

Let us first consider the case where $q_{\text{U(1)}}=0$, either because the HFB PES ${\bold H}^{00}(q; \theta)$ minimizes for $q_{\text{U(1)}}=0$ or because the solution is constrained to it. In this situation, $B_{q \theta}$ is reduced to the particle-number conserving Slater determinant $| \Phi(q;0) \rangle = | \Phi(q) \rangle$ obtained from the constrained HF equations for which $q$ associated with other symmetries than $U(1)$ can still be non zero as briefly described in App.~\ref{SRpartitioning2}. The projectors on ${\cal P}$ and ${\cal Q}$ spaces reduce in that case to
\begin{subequations}
\label{projectorsFockspace_SR}
\begin{align}
{\cal P}(q) &\equiv  | \Phi(q) \rangle \langle \Phi(q) | \, , \label{projectorsFockspace_SR1} \\
{\cal Q}(q) &\equiv  1 - {\cal P}(q) \, , \label{projectorsFockspace_SR2}
\end{align}
\end{subequations}
such that Eq.~\eqref{part_comm} is not fulfilled anymore. Furthermore, the one-body operator $F_{[| \Theta \rangle]}$ is naturally constructed from the SR unperturbed state $| \Theta \rangle = | \Phi(q) \rangle$ such that Eq.~\eqref{F_sym} is also lost along the way. Eventually, either of these two features implies that Eq.~\eqref{H0_comm} is violated as well, i.e. the partitioning becomes symmetry breaking in the SR limit. 

With these elements at hand, the unperturbed Hamiltonian of PGCM-PT becomes
\begin{subequations}
\label{value_E0SR}
\begin{align}
H_0(q) \equiv E^{(0)}(q) {\cal P}(q) + {\cal Q}(q) F_{[| \Phi(q) \rangle]} {\cal Q}(q) \, , \label{def_H02SR}
\end{align}
with 
\begin{align}
E^{(0)}(q) &\equiv \langle \Phi(q) | F_{[| \Phi(q) \rangle]} | \Phi(q) \rangle \nonumber \\
&=\sum_{a_1 b_1} f^{a_1}_{b_1} \, \left[\rho^{\Phi(q)}\right]^{b_1}_{a_1}  \, . 
\end{align}
\end{subequations}

Generally, the above definition of $H_0$ does not match the one at play in MBPT. Only in the unconstrained case, i.e. whenever $\lambda_{q}=0$ in the set of constrained HF equations displayed in App.~\ref{SRpartitioning2}, does the SR reduction of PGCM-PT directly relate to M{\o}ller-Plesset MBPT. In particular, the unperturbed Slater determinant $| \Phi(q) \rangle$ is built from the eigenstates of the one-body operator $F_{[| \Phi(q) \rangle]}$ in that special case while it is not true otherwise. Furthermore, Eq.~\eqref{def_H02SR} becomes
\begin{align}
H_0(q) &= F_{[| \Phi(q) \rangle]} \nonumber \\
&= E^{(0)}(q) + \sum_{k} e_{k}(q) :A^{k}_{k}: \, , \label{def_H0_SR1}
\end{align}
where the latter equality makes use of the one-body eigenbasis of $F_{[| \Phi(q) \rangle]}$ and where 
\begin{align}
E^{(0)}(q) &\equiv \sum_{i=1}^{\text{A}} e_{i}(q) \, .
\end{align}
While the definition of $E^{(0)}(q)$ above is at variance with the choice made in App.~\ref{SRpartitioning2} for M{\o}ller-Plesset MBPT, it only shifts $H_0(q)$ by a constant such that both expansions match from the first order on. Details of the corresponding expansion are discussed in App.~\ref{SRpartitioning2}.

\subsubsection{$U(1)$-breaking single-reference limit}
\label{SR_nosym_lim_part}

In the more general case, the set $B_{q \theta}$ reduces to a particle-number breaking Bogoliubov state $| \Phi(q) \rangle$ in the single-reference limit. Formally, Eqs.~\eqref{projectorsFockspace_SR}-\eqref{value_E0SR} still hold and $H_0$ does not match the unperturbed operator at play in single-reference Bogoliubov many-body perturbation theory (BMBPT)~\cite{Duguet:2015yle,Tichai18BMBPT,Arthuis:2018yoo,Demol:2020mzd,Tichai2020review} (see App.~\ref{SRpartitioning1}). 

However, and contrary to Sec.~\ref{SR_sym_lim_part}, $| \Phi(q) \rangle$ cannot be an eigenstate of the $U(1)$-conserving one-body operator $F$ such that even in the unconstrained case, i.e. whenever $\lambda_{q}=0$, the SR reduction of PGCM-PT does not match M{\o}ller-Plesset BMBPT. Correspondingly, and even though $| \Phi(q) \rangle$ is an eigenstate of $H_0$ by construction, the eigenstates in ${\cal Q}$ space differ from the elementary quasi-particle excitations of $| \Phi(q) \rangle$  (Eq.~\eqref{excitations}) and cannot be directly accessed. As a result, the perturbative expansion is less straightforward to implement than in standard BMBPT where $H_0$ is a generalized, i.e. particle-number-non-conserving, one-body operator whose eigenstates are nothing but $| \Phi(q) \rangle$ and its elementary quasi-particle excitations (see App.~\ref{SRpartitioning1}). 

It is of interest to see to what extent the partitionings at play in (B)MBPT on the one hand and in the SR reduction of PGCM-PT on the other hand do influence numerical results. This comparison is performed in Paper III.

\subsection{Application to second order}
\label{second_order}

Now that the unperturbed reference state and the associated partitioning have been introduced, the perturbative expansion built according to the formal perturbation theory recalled in Sec.~\ref{formal_PT} is specified up to second order, thus defining the PGCM-PT(2) approximation. 

\subsubsection{Zeroth and first-order energies}
\label{0th1storderE}

Given the unperturbed state $| \Theta^{(0)} \rangle \equiv \ket{\mathrm \Theta^{\tilde{\sigma}0}_{\mu} }$ delivered by Eqs.~\eqref{PGCMstate} and~\eqref{HWG_equation}, the zeroth-order energy is given by Eq.~\eqref{value_E0} whereas the first-order energy is obtained through 
\begin{align}
E_{\text{ref}} &= E^{(0)} + E^{(1)}\nonumber \\
&=\langle \Theta^{(0)} | H | \Theta^{(0)} \rangle \nonumber \\
&= \sum_{pq} f^{\ast}(p) H^{\tilde{\sigma}}_{p0q0} \, f(q)   . \label{ref_E_PGCM}
\end{align}

\subsubsection{First-order interacting space}
\label{1storderWF}

According to Eq.~\eqref{EPT2}, the second-order energy $E^{(2)}$ requires the knowledge of the first-order wave-function. Accessing $| \Theta^{(1)} \rangle$ is rendered non-trivial by the fact that  ${\cal Q}$-space eigenstates of $H_0$ are not known a priori. This difficulty leads to the necessity to solve Eq.~\eqref{linearsystemvector} \footnote{The more elaborate Eq.~\eqref{linearsystemmatrix2} needs to be solved to access $| \Theta^{(k)} \rangle$ with $k>1$.}. 

However, solving Eq.~\eqref{linearsystemvector} requires the identification of a suitable basis of ${\cal Q}$ space, i.e. the appropriate {\it first-order interacting space} over which $| \Theta^{(1)} \rangle$ can be exactly expanded. In standard single-reference\footnote{Standard MR perturbation theories rely on an unperturbed state mixing orthogonal elementary excitations of a common vacuum state restricted to a certain valence/active space. In such a situation, the first-order interacting space is also well partitioned~\cite{Tichai:2017rqe} as it is built (in the case of a Hamiltonian containing up to two-body operators) out of single and double excitations outside the valence/active space from each orthogonal product state entering the unperturbed state wave function.} perturbation theories, the first-order wave function is a linear combination of single and double excitations of the unperturbed state, i.e. the first-order interacting space is well partitioned. In the present case, the PGCM unperturbed state prevents a straightforward identification of the first-order interacting space in terms of elementary excitations of a preferred reference vacuum. Indeed, each excitation of a Bogoliubov product state entering $| \Theta^{(0)} \rangle$ can have a non-zero overlap with any of the other HFB vacua making up $| \Theta^{(0)} \rangle$, and thus with $| \Theta^{(0)} \rangle$ itself. Eventually, this means that (i) ${\cal Q}$ cannot be built explicitly and that (ii) Eq.~\eqref{linearsystemvector} cannot be solved exactly. While the first difficulty can be bypassed by using Eq.~\eqref{projectorsFockspace2} repeatedly, the second one requires a procedure to optimally approximate the first-order interacting space.

Rather than referring to the orthonormal representation of ${\cal H}_{\text{A}}$ associated with a preferred reference vacuum and its elementary excitations, one can appropriately consider the multiple representations built out of each product state entering $| \Theta^{(0)} \rangle$, i.e. each Bogoliubov state belonging to  $B_{q \theta}$. This leads to writing the ansatz
\begin{align}
| \Theta^{(1)} \rangle  &\equiv \frac{d_{\tilde{\sigma}}}{v_{\text{G}}}  \sum_q  \sum_\theta \sum_{I} a^{I}(q;\theta) | \Phi^{I}(q;\theta) \rangle \nonumber \\
&= \frac{d_{\tilde{\sigma}}}{v_{\text{G}}}  \sum_q  \sum_\theta \sum_{I} a^{I}(q;\theta) R(\theta) | \Phi^{I}(q) \rangle \, , \label{wf1}
\end{align}
where the index $I$ runs over all singly (S), doubly (D), triply (T)\ldots excitated Bogoliubov vacua $| \Phi^{I}(q;\theta) \rangle$ defined in Eq.~\eqref{rotated_excitations}. The second line of  Eq.~\eqref{wf1} has been obtained thanks to Eq.~\eqref{rotated_excitations2} whereas the coefficients 
\begin{align*}
\{a^{I}(q;\theta); q \in \text{set}  \, \, , \, \, \theta \in D_{\text{G}} \, \, \text{and} \, \, I \in \text{S,D,T,\ldots}\}
\end{align*}
denote the unknowns to be determined. 

\begin{figure*}
    \centering
    \includegraphics[width=0.85\textwidth]{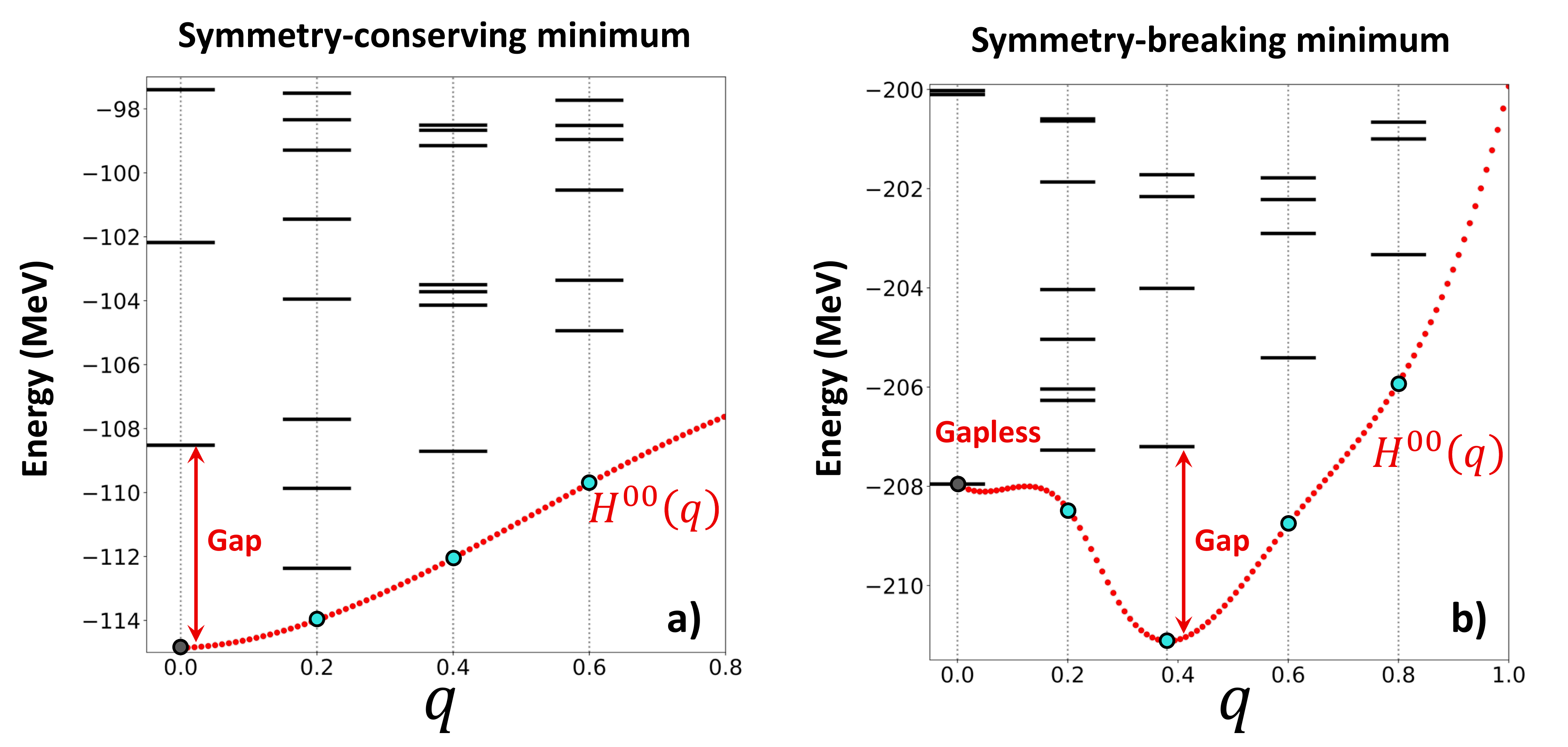}
    \caption{(color online) Schematic energetic representation as a function of the norm $q$ of the order parameter of the considered symmetry of excited Bogoliubov states $\{| \Phi^{I}(q) \rangle; q \in \text{set}  \, \, \text{and} \, \, I \in \text{S,D,T,\ldots}\}$ employed to expand $| \Theta^{(1)} \rangle$. The (red) dotted curve represents the constrained HFB energy  ${\bold H}^{00}(q; 0)$ associated with the vacua $B_q = \{| \Phi(q) \rangle; q \in \text{set} \}$ entering the unperturbed PGCM state $| \Theta^{(0)} \rangle$, and the black bars represent elementary excitations on top of these vacua. Left (right) panel: system characterized by a symmetry-conserving (-breaking) minimum of ${\bold H}^{00}(q; 0)$. In single-reference methods, the problematic gapless symmetry-conserving solution encountered in open-shell nuclei is replaced by a gentle gapful one by allowing the unperturbed state to spontaneously break the symmetry.}
    \label{fig:multiple_rep_HA}
\end{figure*}

The fact that, as pointed out in Sec.~\ref{sym_part}, the first-order wave function is given by
\begin{align}
| \Theta^{(1)} \rangle & \equiv  P^{\tilde{\sigma}}_{00}| \bar{\Theta}^{(1)} \rangle \,\label{1th_order_new}
\end{align}
fully fixes the dependence of these coefficients on the angle $\theta$ of the order parameter. They must display the separable form
\begin{align}
a^{I}(q;\theta) &\equiv a^{I}(q) D_{\text{00}}^{\tilde{\sigma}\ast}(\theta) \, , \label{angle_dep_coeff}
\end{align}
which drastically reduces the cardinality of the set of coefficients to
\begin{align*}
 \{a^{I}(q); q \in \text{set}  \, \, \text{and} \, \, I \in \text{S,D,T,\ldots}\} \, .
\end{align*}
Explicitly projecting onto ${\cal Q}$ space to only retain the orthogonal component to $| \Theta^{(0)} \rangle$, Eq.~\eqref{angle_dep_coeff} is used to rewrite Eq.~\eqref{wf1} under the compact form
\begin{align}
| \Theta^{(1)} \rangle  &= \sum_q \sum_{I} a^{I}(q) | \Omega^{I}(q) \rangle   \, ,\label{wf1_B}
\end{align}
where the expansion now runs over the reduced set of non-orthogonal states
\begin{align}
| \Omega^{I}(q) \rangle  &\equiv {\cal Q} P^{\tilde{\sigma}}_{00} | \Phi^{I}(q) \rangle  \, . \label{basis_states}
\end{align}
As schematically illustrated  in Fig.~\ref{fig:multiple_rep_HA}, the first-order wave-function is thus expanded over (projected) excitations of the HFB vacua carrying different values of the norm $q$ of the order parameter.

In principle, all excitation ranks are involved in Eq.~\eqref{basis_states}, which is unmanageable in practical applications. The idea is to truncate the expansion based on the fact that (i) the Hylleraas functional justifies that an approximation to $| \Theta^{(1)} \rangle$ delivers a variational upper bound to $E^{(2)}$ that can be systematically improved and on the fact that (ii) doing so on the basis of Eq.~\eqref{wf1_B} can provide an optimal approximation. In order to motivate the latter point, let us further investigate the expression of $E^{(2)}$. After noticing that\footnote{Because symmetry blocks associated with different values of $M$ are explicitly separated throughout the whole formalism as explained in Sec.~\ref{sym_part}, ${\cal P} = | \Theta^{(0)} \rangle \langle \Theta^{(0)}|$ is used everywhere in the following.}
\begin{align}
\langle  \Theta^{(0)} | H_1 {\cal Q}   &= \langle  \Theta^{(0)} | (H-H_0) (1-| \Theta^{(0)} \rangle\langle  \Theta^{(0)} |)  \nonumber \\
&= \langle  \Theta^{(0)} |  (H-E_{\text{ref}})  \, , \label{rewritting}
\end{align}
the definition of $\langle  \Theta^{(0)} |$ along with multiple completeness relations in ${\cal H}_{\text{A}}$ are inserted into Eq.~\eqref{EPT2} in order to write the second-order energy as
\begin{align}
E^{(2)} =&  \langle  \Theta^{(0)} | (H-E_{\text{ref}}) | \Theta^{(1)} \rangle  \label{2ndorderE_A} \\ 
=& \frac{d_{\tilde{\sigma}}}{v_{\text{G}}}  \sum_{p\theta} f^{\ast}_{\mu}(p)D_{\text{M0}}^{\tilde{\sigma}}(\theta) \nonumber \\
&\times \!\!\!\sum_{I \in S,D}  \langle \Phi (p;\theta) |  (H\!-\!E_{\text{ref}})  | \Phi^{I}(p;\theta) \rangle \langle \Phi^{I}(p;\theta) | \Theta^{(1)} \rangle  \nonumber
\end{align}
where the excitation rank is naturally truncated given that the two-body Hamiltonian can at most couple each vacuum  $\langle \Phi (p;\theta) |$ to its double excitations. Effectively, Eq.~\eqref{2ndorderE_A} demonstrates that any excited component of $| \Theta^{(1)} \rangle$ in a given representation of ${\cal H}_{\text{A}}$ can only contribute to $E^{(2)}$ if it corresponds to a linear combination of single and double excitations associated with a (possibly) different representation at play. Looking for the first-order interacting space spanned by product states uniquely contributing to $E^{(2)}$, it is thus sufficient to include single and double excitations from each Bogoliubov state entering $| \Theta^{(0)} \rangle$. The approximation presently employed consists thus in replacing Eq.~\eqref{wf1_B} by
\begin{align}
| \Theta^{(1)} \rangle  &= \sum_q \sum_{I \in S,D} a^{I}(q) | \Omega^{I}(q) \rangle   \, . \label{wf1_C}
\end{align}

\subsubsection{Equation of motion}
\label{equation_of_motion}

The last step of the process consists in determining the unknown coefficients $\{a^{I}(q); q \in \text{set}  \, \, \text{and} \, \, I \in \text{S,D}\}$. This is done by solving Eq.~\eqref{linearsystemvector} according to 
\begin{align}
\sum_q \sum_{J \in S,D} M_{IpJq} \, a^{J}(q) &= -  h_1^{I}(p) \, , \label{eq_motion}
\end{align}
with $I \in  S,D$. The ansatz in Eq.~\eqref{wf1_C} does constitute an approximation given that, even if only single and double excitations contribute to the energy, the coefficients are influenced by the presence of higher-rank excitations in the wave function (this is similar to the situation encountered in coupled cluster theory where the energy is a functional of only single and double amplitudes that are themselves influenced by the presence of \emph{higher-rank} amplitudes in the wave-function.). Thus truncating the linear system to singles and doubles defines the working approximation that can be variationally and systematically improved if needed.

The A-body matrix elements entering Eq.~\eqref{eq_motion} are given by
\begin{strip}
\begin{subequations}
\label{ME_linear_PGCMPT2}
\begin{align}
M_{pIqJ} \equiv& \langle \Omega^{I}(p) | H_0-E^{(0)} | \Omega^{J}(q) \rangle \label{ME_linear_PGCMPT2_A} \nonumber  \\
=& \langle \Phi^{I}(p) | {\cal Q}H_0 {\cal Q}P^{\tilde{\sigma}}_{00} | \Phi^{J}(q) \rangle - E^{(0)}\langle \Phi^{I}(p)| {\cal Q} P^{\tilde{\sigma}}_{00}| \Phi^{J}(q) \rangle \nonumber \\
=& F^{\tilde{\sigma}}_{pIqJ} - E^{(0)} N^{\tilde{\sigma}}_{pIqJ} + \sum_{p'q'} f^{\ast}(q')f(p')  \Big( 2E^{(0)} N^{\tilde{\sigma}}_{pIp'0}   N^{\tilde{\sigma}}_{q'0qJ}  - F^{\tilde{\sigma}}_{pIp'0} N^{\tilde{\sigma}}_{q'0qJ} - N^{\tilde{\sigma}}_{pIp'0}F^{\tilde{\sigma}}_{q'0qJ} \Big)\, , \\
h_1^{I}(p) 
\equiv& \langle \Omega^{I}(p) | H_1 | \Theta^{(0)} \rangle \label{ME_linear_PGCMPT2_C} \nonumber  \\
=& \langle \Phi^{I}(p) | (H-E_{\text{ref}}) | \Theta^{(0)} \rangle \nonumber \\
=& \sum_{p'}  \Big(H^{\tilde{\sigma}}_{pIp'0} - E_{\text{ref}} N^{\tilde{\sigma}}_{pIp'0}\Big) f(p') \, , 
\end{align}
\end{subequations}
\end{strip}
where the algebraic expressions of the needed {\it operator kernels}
\begin{align}
O^{\tilde{\sigma}}_{pIqJ} &\equiv  \bra{\Phi^{I}(p)} O P^{\tilde{\sigma}}_{00} \ket{\Phi^{J}(q)} \, , \label{kernels_PGCMPT2}
\end{align}
generalizing those introduced in Sec.~\ref{NOCIeq}, in terms of input quantities are worked out in App.~\ref{PGCMPT2matel}. The kernel associated with the identity operator $N\equiv 1$ is the generalized norm kernel denoted by $N^{\tilde{\sigma}}_{pIqJ}$. In Eq.~\eqref{ME_linear_PGCMPT2}, the sole matrix elements involving excitations in both the ket and the bra are those of a zero or a one-body operator, which limits the complexity of the calculation. Accordingly, matrix elements of the two-body Hamiltonian only involve single or double excitations of the bra.  

\subsubsection{Second-order energy}
\label{2ndorderE}

Once the coefficients $\{a^{I}(q); q \in \text{set}  \, \, \text{and} \, \, I \in \text{S,D}\}$ have been obtained by solving Eq.~\eqref{eq_motion}, the second-order energy can be computed via Eq.~\eqref{EPTbis2}. In expanded form, it reads
\begin{align}
E^{(2)} &= \sum_q \sum_{I \in S,D} h_1^{I\ast}(q) \, a^{I}(q) \nonumber \\
&\equiv \sum_q \sum_{I \in S,D}  e^{(2)I}(q) \nonumber \\
&\equiv \sum_q e^{(2)}_{S}(q) + e^{(2)}_{D}(q) \label{E2_expanded}   \, ,
\end{align}
such that the contributions of each configuration $I$ at deformation $q$ can be isolated, along with their partial sums over the categories of single and double excitations.

One can further compute $E^{(3)}$ from the same information, i.e. from the knowledge of $| \Theta^{(1)} \rangle$. However, according to Eq.~\eqref{EPTbis3}, this requires the evaluation of matrix elements of the two-body Hamiltonian with excited configurations on both sides, which is significantly more costly than the $E^{(2)}$ calculation. These matrix elements are not provided in the present paper but can be worked out to access $E^{(3)}$ in a future paper. 

\begin{figure*}
    \centering
    \includegraphics[width=\textwidth]{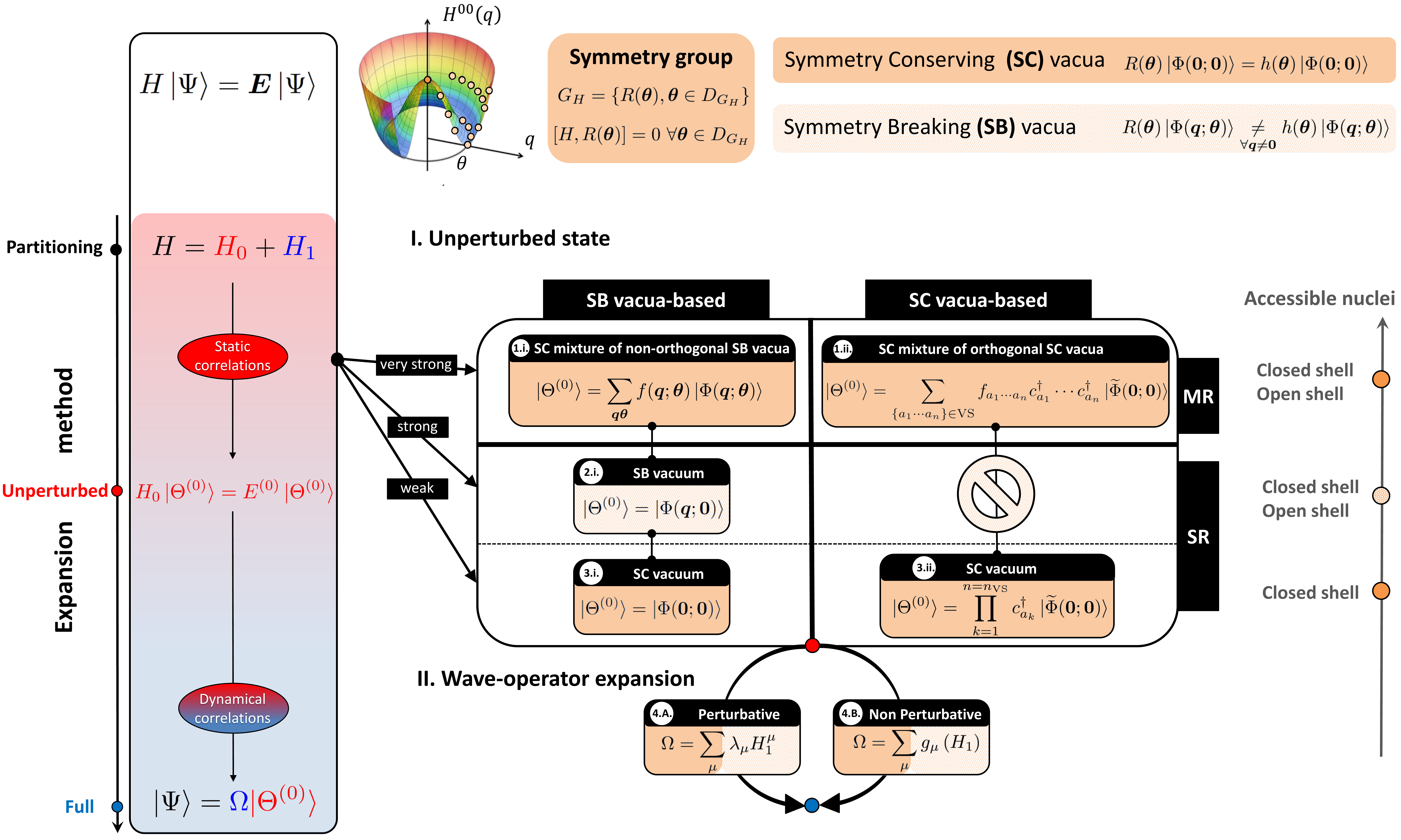}
    \caption{(color online) Schematic representation of the existing options to define the partitioning and unperturbed state at the heart of expansion many-body methods. The left column refers to strategies where one allows vacua $\ket{\Phi(q;\theta)}$ to spontaneously break the symmetries of the Hamiltonian, signalled by a non-zero value for an appropriate order parameter $\boldsymbol{q}\equiv q e^{i\theta}$. 
    Standard single-reference symmetry-conserving schemes (3.i.) appear as particular limits of more general choices, i.e. single-reference symmetry-breaking schemes (2.i.) that are themselves limits of the MR symmetry-conserving scheme (1.i.) introduced in the present work. This is contrasted with a more common philosophy where one only works with a symmetry-conserving vacuum $\ket{\tilde{\Phi}(0;0)}$ corresponding to an appropriately chosen closed-shell core Slater Determinant (right column). Single-reference schemes (3.ii.)  obtained after creating valence nucleons on top of the closed-shell core (VS stands for valence space) appear as particular limits of the symmetry-conserving mixture of symmetry-conserving vacua (1.ii.). For both philosophies, moving up in generality allows one to tackle stronger static correlations, which effectively enlarge the classes of nuclei that can be accessed in a controlled fashion.}
    \label{fig:unperturbed_state}
\end{figure*}

\subsubsection{Linear redundancies}
\label{over_complete}

Linear redundancies pose significant obstacles to the solution of the PGCM-PT(2) linear system (Eq.~\eqref{eq_motion}). While of similar nature as for the PGCM itself, these difficulties are much more acute due to the greater dimensions at play in Eq.~\eqref{eq_motion} compared to the HWG equation (Eq.~\eqref{HWG_equation}) and to the significant linear dependencies of the excitations of the different non-orthogonal HFB vacua. The numerical method implemented to overcome this problem is described in detail in Paper III.

\subsubsection{Intruder states}
\label{intruder_states}

Multi-reference approaches are susceptible to troublesome intruder states that induce singularities in the perturbative corrections. These singularities originate from the possibility that the unperturbed state becomes accidentally degenerate with another eigenstate of $H_0$. This causes the first-order wave-function coefficients to diverge. 

Since it is related to the definition of the partitioning, the nuisance of intruder states can be mitigated by changing the definition of $H_0$. Rather than using an explicitly different $H_0$, the difficulties can be bypassed by adding a constant shift to the chosen $H_0$. While real energy shifts can only move the problematic poles along the real axis~\cite{roos95a}, often causing the divergences to reappear in a slightly different situation, imaginary shifts move the poles into the complex plane and provide a more robust way to remove intruder-state divergences~\cite{forsberg97a}. The corresponding method is detailed in Paper III.

\section{Conclusions}
\label{conclusions}

In the past ten years, perturbative and non-perturbative expansion methods have been instrumental to extend the reach of ab initio calculations over the nuclear chart. Figure~\ref{fig:unperturbed_state} provides a sketched panorama of such many-body methods, detailing in particular their character and their applicability depending on the nature of the unperturbed state and the associated partitioning. While the current limitation to mass numbers $A \lesssim 100$ is primarily computational, there does not exist a symmetry-conserving approach universally applicable to doubly closed-shell, singly open-shell and doubly open-shell nuclei and that scales gently, i.e. polynomially, with the number of nucleons and the single-particle basis size.

The present work addresses such a challenge by formulating a perturbative expansion on top of a MR unperturbed state mixing {\it deformed non-orthogonal Bogoliubov vacua}, i.e. an unperturbed state obtained via the projected generator coordinate method. As a result, (strong) static correlations can be captured in a versatile and efficient fashion at the level of the unperturbed state such that only (weak) dynamical correlations are left to be accounted for via perturbative corrections. Interestingly, the novel method, coined as PGCM-PT, recovers more standard symmetry-conserving and symmetry-breaking single-reference many-body perturbation theories as particular cases.

In addition to being adapted to all types of nuclei, a key feature of PGCM-PT is that it applies to both ground and excited states, i.e. each state coming out of a PGCM calculation can be consistently corrected perturbatively. Another crucial aspect of PGCM-PT relates to the scaling of its cost with nuclear mass. The only other MR perturbation theory applied so far to nuclear systems, MCPT~\cite{Tichai:2017rqe}, is based on an unperturbed state mixing (a large number of) orthogonal Slater determinants built within a limited configuration space out of a symmetry-conserving vacuum\footnote{It can be the particle vacuum whenever the unperturbed state is obtained from a small-scale no-core shell model~\cite{Tichai:2017rqe}.}. 
Contrarily, the PGCM unperturbed state considered here is built from a low-dimensional linear combination of non-orthogonal Bogoliubov product states. 
While it is accessed via a diagonalization procedure, the associated low dimensionality is expected\footnote{This expectation comes from the know-how on PGCM calculations accumulated within the frame of multi-reference energy density functional calculations~\cite{Duguet:2013dga,10.1088/2053-2563/aae0ed}.} to scale much more gently with both mass number and the degree of collective correlations than more traditional MR methods like MCPT. While PGCM-PT corrects PGCM states for dynamical correlations, it is not suited to the description of excited states with a strong "few-elementary excitation" character, \emph{unless} these specific elementary excitations are included into the PGCM ansatz itself. While this is not considered in the present work, such an extension can be naturally envisioned in the future.

The novel PGCM-PT formalism has been laid out in details in the present work. While the generic features of the MR perturbation theory have been described in the bulk of the paper, many technical appendices are provided to fully characterize the approach, in particular the explicit algebraic expressions of the many-body matrix elements constituting the key ingredients to the approach and entering the main equations that need to be solved in practice. The present article is followed by two companion papers in which numerical applications are discussed to characterize the novel method and compare the results to those obtained via existing many-body techniques. 

For the future, it is of interest to envision the possibility to develop a non-perturbative version of PGCM-PT, i.e. a method in which PGCM states are corrected via a non-perturbative expansion.


\newpage
\begin{appendices}

\section{Permutation operators}
\label{permutop}

Many of the algebraic expressions derived below can be economically written via the use of so-called {\it permutation operators} that perform appropriate anti-symmetrizations of the matrix element they act on. A permutation operator $P(s_1/s_2/\ldots/s_n)$, where $s_i$ ($i=\{1,\ldots,n\}$) denotes a given set of indices, permutes the indices belonging to the different sets in all possible ways, {\it without} permuting the indices {\it within} each set. Furthermore, the sign given by the signature of each permutation multiplies the corresponding term. In the present work, the needed permutation operators read as
\begin{strip}
\begin{subequations}
\label{permutoperators}
\begin{align}
    P(k_1/k_2) \equiv{}& 1 - P_{k_1k_2} \, , \\
    P(k_1/k_2k_3) \equiv{}& 1 - P_{k_1k_2} - P_{k_1 k_3}\, , \\
    P(k_1/k_2k_3k_4) \equiv{}& 1 - P_{k_1k_2} - P_{k_1k_3} - P_{k_1k_4} \, , \\
    P(k_1k_2/k_3k_4) \equiv{}& 1 - P_{k_1k_3} 
    - P_{k_1k_4} - P_{k_2k_3} - P_{k_2k_4} + P_{k_1k_3}P_{k_2k_4} \, , \\
    P(k_1/k_2/k_3k_4) \equiv{}& P(k_1k_2/k_3k_4) P(k_1/k_2)
    \nonumber\\
    =& 1 - P_{k_1k_3} 
    - P_{k_1k_4} - P_{k_2k_3} - P_{k_2k_4} + P_{k_1k_3}P_{k_2k_4}+
    P_{k_3k_4} + P_{k_1k_3} P_{k_3k_4}
    \nonumber\\
    &+ P_{k_1k_4}P_{k_3k_4} + P_{k_2k_3}P_{k_3k_4} + P_{k_2k_4}P_{k_3k_4} - P_{k_1k_4}P_{k_2k_3}
    \, , \\
    P(k_1/k_2/k_3/k_4) \equiv{}& P(k_1k_2/k_3k_4) P(k_1/k_2)P(k_3/k_4) \, ,
\end{align}
\end{subequations}
\end{strip}
where the {\it exchange operator} $P_{k_ik_j}$ commutes indices $k_i$ and $k_j$.

\section{Symmetry group}
\label{symgroup}

The symmetry group of $H$ underlines the symmetry quantum numbers carried by its many-body eigenstates. In the present context, the group\footnote{One can add translation and time-reversal symmetries to the presentation to reach the complete symmetry group of the nuclear Hamiltonian.}
\begin{eqnarray}
\text{G}_{H} \equiv \text{SU(2)} \times \text{I} \times \text{U(1)}_N \times \text{U(1)}_Z
\label{group}
\end{eqnarray}
associated with the conservation of total angular momentum, parity and neutron/proton numbers is explicitly considered. The group is a compact Lie group but is non Abelian as a result of SU(2).

\subsection{Unitary representation}

Each subgroup is represented on Fock space $\cal{F}$ via the set of unitary rotation operators
\begin{subequations}
\label{rotation_operators}
\begin{align}
R_{\vec{J}}(\Omega) &\equiv e^{-\imath\alpha J_z} e^{-\imath\beta J_y} e^{-\imath\gamma J_z} \, , \\
R_{N}(\varphi_{n}) &\equiv e^{-i\varphi_{n} N} \, , \\
R_{Z}(\varphi_{p}) &\equiv e^{-i\varphi_{p} Z}  \, , \\
\Pi(\varphi_\pi) &\equiv e^{-i\varphi_{\pi} F} \, ,
\end{align}
\end{subequations}
where $\Omega\equiv(\alpha,\beta,\gamma)$,  $\varphi_\pi$ and $\varphi_{n}$ ($\varphi_{p}$) denote Euler, parity and neutron- (proton-) gauge angles, respectively. The one-body operators entering the unitary representations of interest denote the generators of the group made out of the three components of the total angular momentum $\vec{J} = (J_x,J_y,J_z)$, neutron- (proton-) number $N$ ($Z$) operators as well as of the one-body operator  
\begin{align}
F & \equiv \sum_{ab} f_{ab} c^{\dagger}_a c_b
\end{align}
defined through its matrix elements~\cite{egido91a} 
\begin{align}
f_{ab} & \equiv  \frac{1}{2}\left(1- \pi_a \right) \delta_{ab} \, ,
\end{align}
where $\pi_a$ denotes the parity of one-body basis states that are presently assumed to carry a good parity. The eigenstates of $H$ are characterized by
\begin{subequations}
\label{eigenvalues_symmetry_operators}
\begin{align}
J^2 |\Psi^{\sigma}_{\mu} \rangle &\equiv \hbar^2 J(J+1) |\Psi^{\sigma}_{\mu} \rangle \, , \\
J_z |\Psi^{\sigma}_{\mu} \rangle &\equiv \hbar M |\Psi^{\sigma}_{\mu} \rangle \, , \\
N |\Psi^{\sigma}_{\mu} \rangle &\equiv \text{N} |\Psi^{\sigma}_{\mu} \rangle \, , \\
Z |\Psi^{\sigma}_{\mu} \rangle &\equiv \text{Z} |\Psi^{\sigma}_{\mu} \rangle \, , \\
\Pi(\pi) |\Psi^{\sigma}_{\mu} \rangle &\equiv \Pi |\Psi^{\sigma}_{\mu} \rangle \, ,
\end{align}
\end{subequations}
where $\sigma \equiv (\text{J}\text{M}\Pi\text{NZ})$ and where the operator $\Pi(\pi)$ is nothing but the parity operator and $J^2 \equiv \vec{J} \cdot \vec{J}$ is the Casimir of SU(2). 
  
The irreducible representations (IRREPs) of the group are given by~\cite{VaMo88}
\begin{align}
\langle \Psi^{\sigma}_{\mu} |  R(\theta)  |\Psi^{\sigma'}_{\mu'} \rangle \equiv& D_{\text{MM'}}^{\tilde{\sigma}}(\theta)  \delta_{\tilde{\sigma}\tilde{\sigma}'}  \delta_{\mu\mu'} \, , \label{irreps}
\end{align}
with $\tilde{\sigma} \equiv (\text{J}\Pi\text{NZ})$ and
\begin{align}
D_{\text{MM'}}^{\tilde{\sigma}}(\theta) \equiv& D_{\text{MM'}}^{\text{J}}(\Omega) e^{-i\varphi_{n} \text{N}}e^{-i\varphi_{p} \text{Z}} e^{-\frac{i}{2}(1-\Pi)\varphi_\pi}  \, , \label{irreps2}
\end{align}
and where the rotation operators have been gathered into
\begin{equation}
R(\theta) \equiv R_{\vec{J}}(\Omega)R_{N}(\varphi_{n})R_{Z}(\varphi_{p})\Pi(\varphi_\pi) \, ,
\label{rotation_operator}
\end{equation}
with
\begin{align}
\theta\equiv(\Omega, \varphi_n,\varphi_p, \varphi_\pi)
\end{align}
encompassing all rotation angles. The domain of definition of the group is thus
\begin{align}
D_{\text{G}_H} & \equiv D_{\alpha} \times D_{\beta} \times D_{\gamma} \times  D_{\varphi_{n}}  \times  D_{\varphi_{p}}  \times  D_{\varphi_{\pi}}  \\
&= [0,4\pi]\times[0,\pi]\times[0,2\pi]\times[0,2\pi]\times[0,2\pi]\times\{0,\pi\} \nonumber \, .
\end{align}
In Eq.~\eqref{irreps}, $D^{\text{J}}_{\text{MM'}}(\Omega)$ denotes Wigner D-matrices that can be expressed in terms of (real) reduced Wigner $d$-functions through $D^{\text{J}}_{\text{MM'}}(\Omega)\equiv e^{-i\text{M}\alpha} \, d^{\text{J}}_{\text{MM'}}(\beta) \, e^{-i\text{M'}\gamma}$. 

Given that the degeneracy of the IRREPs is $d_{\tilde{\sigma}} = 2J+1$ and the volume of the group is
\begin{align}
v_{\text{G}_H} &\equiv \int_{D_{\text{G}}} \text{d}\theta \nonumber \\
&\equiv \sum_{\varphi_\pi=0,\pi} \int_{[0,4\pi]\times[0,\pi]\times[0,2\pi]} \hspace{-2.1cm}d\alpha \sin \beta \, d\beta \, d\gamma \, \int_{0}^{2\pi} \!\! \text{d}\varphi_{n}\int_{0}^{2\pi} \!\!\text{d}\varphi_{p}  \nonumber \\
&= 2(16\pi^2)(2\pi)^2 \, 
\end{align}
the orthogonality of the IRREPs read as
\begin{align}
 \int_{D_{\text{G}}} \text{d}\theta  D_{\text{MK}}^{\tilde{\sigma}\ast}(\theta) D_{\text{M'K'}}^{\tilde{\sigma}'}(\theta) &= \frac{v_{\text{G}_H}}{d_{\tilde{\sigma}}} \delta_{\tilde{\sigma}\tilde{\sigma}'} \delta_{MM'} \delta_{KK'} \, . 
\end{align}
Furthermore, the unitarity of the symmetry transformations, i.e. $R^{\dagger}(\theta)R(\theta)=R(\theta)R^{\dagger}(\theta)=1$, induces
\begin{subequations}
\label{unitarity}
\begin{align}
\sum_{M} D_{\text{MK}}^{\tilde{\sigma}\ast}(\theta) D_{\text{MK'}}^{\tilde{\sigma}}(\theta) &=  \delta_{KK'} \, , \label{unitarity1} \\
\sum_{K}  D_{\text{MK}}^{\tilde{\sigma}}(\theta)D_{\text{M'K}}^{\tilde{\sigma}\ast}(\theta) &= \delta_{MM'} \, . \label{unitarity2}
\end{align}
\end{subequations}
An irreducible tensor operator $T^{\tilde{\sigma}}_{K}$ of rank $J$ and a state $| \Psi^{\tilde{\sigma}K}_{\mu} \rangle$ transform under rotation according to
\begin{subequations}
\label{eq:ten:def}
\begin{eqnarray}
R(\theta) \, T^{\tilde{\sigma}}_{K} \, R(\theta)^{-1}  &=& \sum_{M}  T^{\tilde{\sigma}}_{M} \, D^{\tilde{\sigma}}_{MK}(\theta) \,\,\, ,\label{eq:ten:def1} \\
R(\theta) \, | \Psi^{\tilde{\sigma} K}_{\mu} \rangle &=& \sum_{M}  | \Psi^{\tilde{\sigma} M}_{\mu} \rangle \, D^{\tilde{\sigma}}_{MK}(\theta) \,\,\, . \label{eq:ten:def2}
\end{eqnarray}
\end{subequations}

Peter-Weyl's theorem ensures that any function $f(\theta) \in L^2(\text{G}_H)$ can be expanded according to
\begin{equation}
f(\theta) \equiv \sum_{\tilde{\sigma}} \sum_{MK} \, f^{\tilde{\sigma}}_{MK} \, \, D^{\tilde{\sigma}\ast}_{MK}(\theta) \, , \label{decomposition_general}
\end{equation}
such that the set of complex expansion coefficients $\{f^{\tilde{\sigma}}_{MK}\}$ can be extracted thanks to the orthogonality of the IRREPs through
\begin{equation}
f^{\tilde{\sigma}}_{MK} = \frac{d_{\tilde{\sigma}}}{v_{\text{G}_H}} \int_{D_{\text{G}_H}} \text{d}\theta  D_{\text{MK}}^{\tilde{\sigma}}(\theta) f(\theta) \, . \label{extractcoeff}
\end{equation}

\subsection{Projection operators}

The operator
\begin{align}
P^{\sigma} &\equiv P^{\text{J}}_{\text{M}} P^{\text{N}} P^{\text{Z}} P^{\Pi} \, \label{collecprojector}
\end{align}
collects the projection operators on good symmetry quantum numbers
\begin{subequations}
\label{projector}
\begin{align}
P^{\text{J}}_{\text{M}} &\equiv \sum_K g_K  P^{\text{J}}_{\text{MK}}  \nonumber \label{projectorJ} \\
&\equiv \sum_K g_K   \frac{2J+1}{16\pi^2}\int_{[0,4\pi]\times[0,\pi]\times[0,2\pi]} \hspace{-2.1cm}\ud \Omega \, D^{\text{J}*}_{\text{MK}}(\Omega) R_{\vec{J}}(\Omega)   \, ,  \\
P^{\text{N}}  &\equiv  \frac{1}{2\pi}\int_{0}^{2\pi} \ud \varphi_{n} e^{i\varphi_{n} \text{N}} R_{N}(\varphi_{n}) \, , \label{projectorN} \\
P^{\text{Z}}  &\equiv   \frac{1}{2\pi}\int_{0}^{2\pi} \ud \varphi_{p} e^{i\varphi_{p} \text{Z}} R_{Z}(\varphi_{p})  \, , \label{projectorZ} \\
P^{\Pi}  &\equiv   \frac{1}{2} \sum_{\varphi_\pi=0,\pi} e^{\frac{i}{2}(1-\Pi)\varphi_\pi} \Pi(\varphi_\pi)\, , \label{projectorPi} 
\end{align}
\end{subequations}
such that one can write in a compact way
\begin{align}
P^{\sigma} &=\, \frac{d_{\tilde{\sigma}}}{v_{\text{G}}} \sum_K g_K  \int_{D_{\text{G}}} \text{d}\theta  D_{\text{MK}}^{\tilde{\sigma}\ast}(\theta) R(\theta) \nonumber \\
&\equiv \sum_K g_K  P^{\tilde{\sigma}}_{\text{MK}} \, . \label{collecprojector2}
\end{align}
The so-called {\it transfer operator} $P^{\tilde{\sigma}}_{\text{MK}}$ fulfills
\begin{subequations}
\begin{align}
P^{\tilde{\sigma}}_{\text{MK}} &=  \sum_{\mu} | \Psi^{\tilde{\sigma}M}_{\mu} \rangle \langle \Psi^{\tilde{\sigma}K}_{\mu} | \, , \label{transferop2} \\
P^{\tilde{\sigma}\dagger}_{\text{MK}} &= P^{\tilde{\sigma}}_{\text{KM}}   \, , \label{transferop1} \\
P^{\tilde{\sigma}}_{\text{MK}} P^{\tilde{\sigma}'}_{\text{M'K'}} &= \delta_{\tilde{\sigma}\tilde{\sigma}'}\delta_{KM'} P^{\tilde{\sigma}}_{\text{MK'}} \, ,
\end{align}
\end{subequations}
along with the identity
\begin{align}
P^{\sigma} R(\theta) &= \sum_K g_K \sum_{M'} D_{\text{KM'}}^{\tilde{\sigma}}(\theta) P^{\tilde{\sigma}}_{\text{MM'}}  \, .\label{projsurrot}
\end{align}
The present paper is eventually interested in the particular case where $g_K = \delta_{K0}$.

\section{Bogoliubov algebra}
\label{bogoalgebra}

\subsection{Operators definition}
\label{operators}

Given a basis $\mathcal B_1 \equiv \{| l \rangle\}$ of the one-body Hilbert space \(\mathcal H_1\) whose associated set of particle creation and annihilation operators is denoted as \( \{c_l^\dagger, c_l\}\), an arbitrary particle-number-conserving operator \(O\) is represented as
\begin{equation} 
	O\equiv\sum_{n=0}^{r} {O^{nn}} \, , \label{generalop}
\end{equation}
where each \(n\)-body component\footnote{The term $O^{00}$ is a number.}
\begin{equation} 
	O^{nn}\equiv \frac{1}{n!} \frac{1}{n!}  
	\sum_{\substack{a_1\cdots a_n\\b_1\cdots b_n}}
	o^{a_1\cdots a_n}_{b_1\cdots b_n} \, 
	C^{a_1\cdots a_n}_{b_1\cdots b_n} \, . \label{defnbodyopvacuum}
\end{equation}
Here
\begin{equation} 
C^{a_1\cdots a_n}_{b_1\cdots b_n}\equiv c^\dag_{a_1}\cdots c^\dag_{a_n}c_{b_n}\cdots c_{b_1} \label{stringpart}
\end{equation}
defines a string of $n$ particle creation and $n$ particle annihilation operators such that 
\begin{equation}
\left(C^{a_1\cdots a_n}_{b_1\cdots b_n}\right)^{\dagger}=C_{a_1\cdots a_n}^{b_1\cdots b_n} \, .
\end{equation}
The string is in normal order with respect to the particle vacuum \(\ket 0\) 
\begin{equation} 
N(C^{a_1\cdots a_n}_{b_1\cdots b_n})=C^{a_1\cdots a_n}_{b_1\cdots b_n} \, ,
\end{equation}
where $N(\ldots)$ denotes the normal ordering with respect to \(\ket 0\), and it is anti-symmetric under the exchange of any pair of upper or lower indices, i.e.
\begin{equation}
  C^{a_1\cdots a_n}_{b_1\cdots b_n} = \epsilon(\sigma_u) \epsilon(\sigma_l)  \, C^{\sigma_u(a_1\cdots a_n)}_{\sigma_l(b_1\cdots b_n)} \, ,
\end{equation}
where $\epsilon(\sigma_u)$  ($\epsilon(\sigma_l)$) refers to the signature of the permutation $\sigma_u(\ldots)$ ($\sigma_l(\ldots)$) of the $n$ upper (lower) indices. 

In Eq.~\eqref{defnbodyopvacuum}, the $n$-body matrix elements $\{o^{a_1\cdots a_n}_{b_1\cdots b_n}\}$ constitute a mode-$2n$ tensor, i.e. a data array carrying $2n$ indices, associated with the string they multiply. The $n$-body matrix elements are also fully anti-symmetric under the exchange of any pair of upper or lower indices, i.e.
\begin{equation}
  o^{a_1\cdots a_n}_{b_1\cdots b_n} = \epsilon(\sigma_u) \epsilon(\sigma_l)  \, o^{\sigma_u(a_1\cdots a_n)}_{\sigma_l(b_1\cdots b_n)} \, .
\end{equation}

\subsection{Bogoliubov state}
\label{bogovacuum}

The linear Bogoliubov transformation \(\mathcal W(q)\) connects the set of particle creation and annihilation operators to a set of quasi-particle creation and annihilation operators obeying fermionic anti-commutation rules
\begin{subequations}
\label{anticommqp}
\begin{align}
  \{\beta_k(q),\beta_l(q)\} &= 0 \, , \\
  \{\beta^\dagger_k(q), \beta^\dagger_l(q)\} &= 0 \, , \\
  \{\beta_k(q), \beta^\dagger_l(q)\} &= \delta_{kl} \, .
\end{align}
\end{subequations}
Formally the transformation reads~\cite{RiSc80}
\begin{equation}
  \begin{pmatrix}
\beta(q)\\ \beta^\dagger(q)
  \end{pmatrix}
  \equiv \mathcal W^\dagger(q)
  \begin{pmatrix}
c\\ c^\dagger
  \end{pmatrix} \, , \label{transfobogo1}
\end{equation}
with
\begin{equation}\label{eq:bogo-trans}
\mathcal W(q)\equiv
  \begin{pmatrix}
    U(q)&V^*(q)\\
    V(q)&U^*(q)
  \end{pmatrix} \, ,
\end{equation}
so that the Bogoliubov transformation in expanded form reads as
\begin{subequations}
\label{transfobogo2}
\begin{eqnarray}
  \beta_k(q) &\equiv& \sum_l {U^*}_{lk}(q) \, c_l + {V^*}_{k}^{l}(q) \, c^\dagger_l \,,\\
  \beta_k^\dagger(q) &\equiv& \sum_l  U^{lk}(q) \, c^\dagger_l + V^{k}_l(q) \ c_l \,.
\end{eqnarray}
\end{subequations}
The anti-commutation rules (Eq.~\eqref{anticommqp}) constrain \(\mathcal W(q)\) to be unitary
\begin{equation}
  \mathcal W^\dagger(q) \mathcal W(q) = \mathcal W(q) \mathcal W^\dagger(q) = 1 \, ,
\end{equation}
which translates into
\begin{subequations}
\label{unitaryBogo}
  \begin{eqnarray}
    U^\dagger(q) U(q) + V^\dagger(q) V(q) = & 1 \, , \label{unitaryBogoA} \\
    V^T(q) U(q) + U^T(q) V(q)        = & 0 \, , \label{unitaryBogoB} \\
    U(q) U^\dagger(q) +  V^*(q) V^T(q)   = & 1 \, , \label{unitaryBogoC} \\
    V(q) U^\dagger(q) + U^*(q) V^T(q)    = & 0 \, . \label{unitaryBogoD}
  \end{eqnarray}
\end{subequations}

The normalized Bogoliubov product state \(| \Phi(q) \rangle\) is defined, up to a phase, as the vacuum of the  quasi-particle operators, i.e. 
\begin{equation}
\beta_k(q) | \Phi(q) \rangle \equiv 0 \,, \ \forall k \, .
\end{equation} 
Contrary to Slater determinants, which constitute a subset of Bogoliubov states, the latter are not eigenstates of neutron and proton number operators in general.

\subsection{One-body density matrices}
\label{HFBdensitymatrices}

The Bogoliubov vacuum is fully characterized by its normal $\rho(q)$ and anomalous $\kappa(q)$ one-body density matrices whose matrix elements are defined through
\begin{subequations}
  \label{eq:normal_anomalous_densities}
  \begin{align}
    \rho^{l_1}_{l_2}(q)
    &\equiv \langle \Phi(q) | c^\dagger_{l_2} c_{l_1} | \Phi(q) \rangle = V^\ast(q) V^T(q) \, , \\*
    \kappa^{l_1l_2}(q)
    &\equiv \langle \Phi(q) | c_{l_2} c_{l_1} | \Phi(q) \rangle = V^\ast(q) U^T(q) \, ,
  \end{align}
\end{subequations}
such that 
\begin{subequations}
  \begin{align}
    \rho^\dagger(q)
    &= \rho(q) \, , \\*
    \kappa^T(q)
    &= - \kappa(q) \, .
  \end{align}
\end{subequations}

\subsection{Normal-ordered operators}
\label{normalordererop}

It is eventually useful to normal order operators with respect to the Bogoliubov vacuum $| \Phi(q) \rangle$ and express them in terms of the associated quasi-particle operators. Applying standard Wick's theorem leads thus to the rewriting of $O$ according to
\begin{align}
  O&\equiv \sum_{n=0}^{r} {\bold O}^{[2n]}(q) \equiv \sum_{n=0}^{r} \sum_{\substack{i,j=0/\\i+j=2n}}^{2r} {\bold O}^{ij}(q)
  \, , \label{normalorderedopbogo}
\end{align}
with\footnote{The term ${\bold O}^{00}(q)$ is just a number.}
\begin{equation} 
	{\bold O}^{ij}(q)\equiv \frac{1}{i!} \frac{1}{j!}  
	\sum_{\substack{k_1\cdots k_i\\l_1\cdots l_j}}
	{\bold o}^{k_1\cdots k_i}_{l_1\cdots l_j}(q) \, 
	B^{k_1\cdots k_i}_{l_1\cdots l_j}(q) \, , \label{defnbodyopbogovacuum}
\end{equation}
where
\begin{equation} 
B^{k_1\cdots k_i}_{l_1\cdots l_j}(q)\equiv \beta^\dagger_{k_1}(q)\cdots \beta^\dagger_{k_i}(q) \beta_{l_j}(q)\cdots \beta_{l_1}(q) \label{stringQP}
\end{equation}
denotes a string of $i$ quasi-particle creation and $j$ quasi-particle annihilation operators such that 
\begin{equation} 
\left(B^{k_1\cdots k_i}_{l_1\cdots l_j}(q)\right)^{\dagger}=B_{k_1\cdots k_i}^{l_1\cdots l_j}(q) \, .
\end{equation}
The string is in normal order with respect to the Bogoliubov state $| \Phi(q) \rangle$
\begin{equation} 
:B^{k_1\cdots k_i}_{l_1\cdots l_j}(q):=B^{k_1\cdots k_i}_{l_1\cdots l_j}(q) \, ,
\end{equation}
where $:\ldots:$ denotes the normal ordering with respect to $| \Phi(q) \rangle$, and it is anti-symmetric under the exchange of any pair of upper or lower indices, i.e.
\begin{equation}
  B^{k_1\cdots k_i}_{l_1\cdots l_j}(q) = \epsilon(\sigma_u) \epsilon(\sigma_l)  \, B^{\sigma_u(k_1\cdots k_i)}_{\sigma_l(l_1\cdots l_j)}(q) \, .
\end{equation}

In Eq.~\eqref{defnbodyopbogovacuum}, the matrix elements $\{{\bold o}^{k_1\cdots k_i}_{l_1\cdots l_j}(q)\}$ are fully anti-symmetric under the exchange of any pair of upper or lower indices, i.e.
\begin{equation}
  {\bold o}^{k_1\cdots k_i}_{l_1\cdots l_j}(q) = \epsilon(\sigma_u) \epsilon(\sigma_l)  \, {\bold o}^{\sigma_u(k_1\cdots k_i)}_{\sigma_l(l_1\cdots l_j)}(q) \, ,
\end{equation}
and are functionals of the Bogoliubov matrices $(U(q),V(q))$ and of the matrix elements $\{o^{a_1\cdots a_n}_{b_1\cdots b_n}\}$ initially defining the operator $O$. As such, the content of each operator ${\bold O}^{ij}(q)$ depends on the rank r of $O$. For more details about the normal ordering procedure and for explicit expressions of the matrix elements up to $r=3$, see Refs.~\cite{Duguet:2015yle,Tichai18BMBPT,Arthuis:2018yoo,Demol:2020mzd,Tichai2020review,Ripoche2020}.

\subsection{Constrained Hartree-Fock-Bogoliubov theory}
\label{HFBequations}

The state $\ket{\Phi(q)}$ is obtained by minimizing its total energy under the constraints that it satisfies\footnote{In our discussion A stands for either the neutron (N) or the proton (Z) number.}
\begin{subequations}
\label{constraintsHFB}
\begin{align}
\langle \Phi(q) | A | \Phi(q) \rangle &= \text{A} \, , \label{constraintZ} \\
\langle \Phi(q) | Q | \Phi(q) \rangle &= q \, , \label{constraintQ}
\end{align}
\end{subequations}
where $Q$ is a generic operator of interest. To do so, one considers the Routhian
\begin{subequations}
\begin{align}
R &\equiv H - \lambda_{\text{A}} (A-\text{A}) - \lambda_{q} (Q-q) \\
&\equiv \Omega - \lambda_{q} (Q-q) \, , \label{routhian}
\end{align}
\end{subequations}
where $\lambda_{\text{A}}$ and $\lambda_{q}$ denote two Lagrange parameters\footnote{In actual applications, one Lagrange multiplier relates to constraining the neutron number N and one Lagrange multiplier is used to constrain the proton number Z.}. The Routhian reduces to the so-called grand potential $\Omega$ whenever $\lambda_{q}=0$, i.e. when performing unconstrained calculations with respect to the order parameter $q$. Minimizing\footnote{As alluded to in Sec.~\ref{normalordererop}, the explicit functional form of ${\bold R}^{00}(q)$ depends on the initial rank of $H$ and $Q$ and can be found elsewhere for up to 3-body operators~\cite{Duguet:2015yle,Arthuis:2018yoo,Ripoche2020}.}
\begin{align}
R(q) &\equiv \langle \Phi(q) | R | \Phi(q) \rangle= {\bold R}^{00}(q)   \label{HFBrouthian}
\end{align}
according to Ritz' variational principle, the Bogoliubov matrices $(U(q),V(q))$ are found as solutions of the constrained Hartree-Fock-Bogoliubov (HFB) eigenequation~\cite{RiSc80}
\begin{align}
  \label{eq:hfb_equation}
  \begin{pmatrix} \bar{h}(q)  & \bar{\Delta}(q) \\ -\bar{\Delta}^\ast(q) & -\bar{h}^\ast(q) \end{pmatrix} \begin{pmatrix} U(q) \\ V(q) \end{pmatrix}_k
  &= E_k(q) \begin{pmatrix} U(q) \\ V(q) \end{pmatrix}_k \, ,
\end{align}
where the eigenvalues $\{E_k(q)\}$ are referred to as quasi-particle energies. The constrained HFB Hamiltonian matrix 
\begin{align}
\label{eq:hfb_matrix}
{\cal H}(q) &\equiv  \begin{pmatrix} \bar{h}(q)  & \bar{\Delta}(q) \\ -\bar{\Delta}^\ast(q) & -\bar{h}^\ast(q) \end{pmatrix}
   \, ,
\end{align}
is built out of the constrained one-body Hartree-Fock and Bogoliubov fields
\begin{subequations}
\label{constrainedfields}
\begin{align}
\bar{h}^{l}_{l'}(q) 
&\equiv \frac{\partial {\bold R}^{00}(q)}{\partial {\rho^*}^{l}_{l'}(q)} \nonumber \\
&= \langle \Phi(q) |\{[c_l,R],c^{\dagger}_{l'}\}| \Phi(q) \rangle \nonumber \\
&= \frac{\partial {\bold H}^{00}(q)}{\partial {\rho^*}^{l}_{l'}(q)} - \lambda_{q} \frac{\partial {\bold Q}^{00}(q)}{\partial {\rho^*}^{l}_{l'}(q)} - \lambda_{\text{A}} \delta_{ll'} \, , \\
\bar{\Delta}_{ll'}(q) 
&\equiv \frac{\partial {\bold R}^{00}(q)}{\partial \kappa^*_{ll'}(q)} \nonumber \\
&= \langle \Phi(q) |\{[c_l,R],c_{l'}\}| \Phi(q) \rangle  \nonumber \\
&= \frac{\partial {\bold H}^{00}(q)}{\partial \kappa^*_{ll'}(q)} - \lambda_{q} \frac{\partial {\bold Q}^{00}(q)}{\partial \kappa^*_{ll'}(q)} \, , 
\end{align}
\end{subequations}
where 
\begin{align}
\frac{\partial {\bold H}^{00}(q)}{\partial {\rho^*}^{l}_{l'}(q)} &= f^{l}_{l'}[| \Phi(q) \rangle]
\end{align}
delivers nothing but the matrix elements of the one-body operator defined in Eq.~\eqref{def_F} computed from the normal one-body density matrix of $| \Phi(q) \rangle$.

At convergence, where the constraints are satisfied, the HFB energy is
\begin{align}
\langle \Phi(q) | H | \Phi(q) \rangle &= {\bold H}^{00}(q)  = {\bold R}^{00}(q) \, . \label{HFBenergy}
\end{align}
Furthermore, Eq.~\eqref{eq:hfb_equation} implies that 
\begin{align}
\mathcal W^\dagger(q)  {\cal H}(q) \mathcal  W(q)
  &= \begin{pmatrix} {\bold R}^{11}(q)  & {\bold R}^{20}(q)  \\ 
  -{\bold R}^{20\ast}(q) & -{\bold R}^{11\ast}(q) \end{pmatrix} \nonumber
 \\
 &= \begin{pmatrix} E(q)  & 0 \\ 0 & -E(q) \end{pmatrix}\, ,
\end{align}
such that the properties
\begin{subequations}
\label{canonicalR}
\begin{align}
{\bold R}^{20}(q) &= {\bold R}^{02}(q)=0 \, , \\
{\bold R}^{11}(q) &= \sum_{k} E_{k}(q) \beta^{\dagger}_k(q) \beta_k(q)  \, , 
\end{align}
\end{subequations}
are fulfilled at convergence.
  
\subsection{Elementary excitations}
\label{elementaryexcitations}

Given the Bogoliubov state $| \Phi(q) \rangle$, a complete basis of Fock space ${\cal F}$ is obtained by generating all its elementary excitations 
\begin{equation} 
| \Phi^{k_1\cdots k_i}(q) \rangle \equiv B^{k_1\cdots k_i}(q) | \Phi(q) \rangle \, , \label{excitations}
\end{equation}
where $B^{k_1\cdots k_i}(q)$ defines the subclass of strings defined in Eq.~\eqref{stringQP} containing quasi-particle creation operators only.

It is interesting to note that each state defined through Eq.~\eqref{excitations} is itself a Bogoliubov vacuum whose associated Bogoliubov transformation can be deduced from the one defining $| \Phi(q) \rangle$ (Eq.~\eqref{transfobogo2}). Writing as $K_{n}\equiv \{k_1\cdots k_n\}$ the n-tuple defining a given elementary excitation $| \Phi^{K_n}(q) \rangle$, the associated Bogoliubov transformation $(U(q,K_n),V(q,K_n))$ is given by the matrices
\begin{subequations}
\begin{align}
U^{lk}(q,K_n)&\equiv U^{lk}(q) \,\,\,\, \text{if} \,\,\,\, k \notin K_n \, , \\
U^{lk}(q,K_n)&\equiv V^{k*}_{l}(q) \,\,\,\, \text{if} \,\,\,\, k \in K_n \, , \\
V^{l}_{k}(q,K_n)&\equiv V^{l}_{k}(q) \,\,\,\, \text{if} \,\,\,\, k \notin K_n \, , \\
V^{l}_{k}(q,K_n)&\equiv U^*_{lk}(q) \,\,\,\, \text{if} \,\,\,\, k \in K_n \, .
\end{align}
\label{redefexcitvacua}
\end{subequations}
Such a consideration can be exploited to eventually compute matrix elements of operators between two Bogoliubov states that may differ not only by the value of the collective coordinate $q$ but also by the elementary excitation character. The idea of evaluating matrix elements by redefining each elementary excitation of an original Bogoliubov vacuum as a novel vacuum is a generalization of the so-called generalized Slater-Condon rules~\cite{mayer03a}. The present work follows a numerically more efficient route where quasi-particle excitation operators are explicitly processed in order to limit the number of reference vacua to those entering the PGCM unperturbed state $\ket{\mathrm \Theta^{\sigma}_{\mu} }$ introduced in Eq.~\eqref{PGCMstate} and avoid the combinatorics associated with the redefinition of many Bogoliubov transformations through Eq.~\eqref{redefexcitvacua}. 

\subsection{Rotated Bogoliubov state}
\label{rotatedbogovacuum}

Given the Bogoliubov state $| \Phi(q) \rangle$, its rotated partner 
\begin{equation}
  \ket{\Phi(q;\theta)} \equiv R(\theta) \ket {\Phi(q) } \, ,
\end{equation}
is also a Bogoliubov state whose associated quasi-particle operators $\{\beta(q;\theta), \beta^\dagger(q;\theta)\}$ are characterized by the Bogoliubov transformation
\begin{align}
  \mathcal W(q;\theta) &=
  \begin{pmatrix}
    r(\theta)&0\\0&r(\theta)^\dagger
  \end{pmatrix}
  \mathcal W(q)  \nonumber \\
  &\equiv   
  \begin{pmatrix}
    U(q;\theta)&V^*(q;\theta)\\
    V(q;\theta)&U^*(q;\theta)
  \end{pmatrix} 
\, ,
\end{align}
where \(r(\theta)\) defines the matrix representation of \(R(\theta)\) in the one-body Hilbert-space. Its elements are
\begin{equation}
r^{l_1}_{l_2}(\theta) \equiv \langle l_1 | R(\theta) | l_2 \rangle \, .
\end{equation}
Given that $\mathcal W(q;\theta)$ is a unitary Bogoliubov transformation, Eqs.~\eqref{unitaryBogo} is also satisfied when substituting $(U(q),V(q))$ for $(U(q;\theta),V(q;\theta))$.

Because $R(\theta) \in \text{G}_{H}$, the energy of the rotated HFB state
\begin{align}
\langle \Phi(q;\theta) | H | \Phi(q;\theta) \rangle &= {\bold H}^{00}(q;\theta) \,  \label{rotatedHFBenergy}
\end{align}
is in fact independent of the rotation angle; i.e. ${\bold H}^{00}(q;\theta)={\bold H}^{00}(q)$ for all $\theta$.

Elementary excitations of the rotated Bogoliubov state are given by 
\begin{equation} 
| \Phi^{k_1\cdots k_i}(q;\theta) \rangle \equiv B^{k_1\cdots k_i}(q;\theta) | \Phi(q; \theta) \rangle \,  \label{rotated_excitations}
\end{equation}
where the rotated string reads as
\begin{align} 
B^{k_1\cdots k_i}(q;\theta) &\equiv R(\theta) B^{k_1\cdots k_i}(q) R^{\dagger}(\theta) \label{stringrotatedQP}\\
&= \beta^\dagger_{k_1}(q;\theta)\cdots \beta^\dagger_{k_i}(q;\theta) \nonumber \, ,
\end{align}
such that they are nothing but the rotated elementary excitations
\begin{equation} 
| \Phi^{k_1\cdots k_i}(q;\theta) \rangle \equiv R(\theta) | \Phi^{k_1\cdots k_i}(q) \rangle \,  \label{rotated_excitations2} .
\end{equation}

\subsection{Single-reference partitioning}
\label{SRpartitioning}

The presently developed perturbation theory is of MR character due to the fact that the PGCM unperturbed state  (Eq.~\eqref{PGCMstate}) is a linear combination of several Bogoliubov vacua. However, in the limit where the PGCM state reduces to a single Bogoliubov state, which itself reduces to a Slater determinant whenever $U(1)$ symmetry is conserved, PGCM-PT becomes of single-reference character and must thus entertain some connection with single-reference (B)MBPT~\cite{Tichai:2020dna}. To clarify this connection, the partitioning at play in the latter approaches are now briefly recalled.

\subsubsection{BMBPT}
\label{SRpartitioning1}

Because of the inherent necessity to control the average particle number, the operator driving the perturbation in BMBPT is the grand potential $\Omega$~\cite{Tichai:2018vjc}. Whenever $| \Phi(q) \rangle$ results from a {\it constrained} HFB calculation (see Sec.~\ref{HFBequations}), a natural partitioning is given by
\begin{equation}
\label{split1}
\Omega = \Omega_{0}(q) + \Omega_{1}(q) \ ,
\end{equation}
such that
\begin{subequations}
\label{split2}
\begin{align}
\Omega_{0}(q) &\equiv {\bold \Omega}^{00}(q) + \bar{{\bold \Omega}}^{11}(q) \ , \\
\Omega_{1}(q) &\equiv {\bold \Omega}^{20}(q) + \breve{{\bold \Omega}}^{11}(q) + {\bold \Omega}^{02}(q)  \label{e:perturbation}  \\
  &\phantom{\equiv } + {\bold \Omega}^{40}(q) + {\bold \Omega}^{31}(q) + {\bold \Omega}^{22}(q) +  {\bold \Omega}^{13}(q) + {\bold \Omega}^{04}(q)  \ , \nonumber
\end{align}
\end{subequations}
with $\breve{{\bold \Omega}}^{11}(q)\equiv {\bold \Omega}^{11}(q) - \bar{{\bold \Omega}}^{11}(q)$ and where the diagonal one-body part of $\Omega_{0}(q)$ 
\begin{equation}
\bar{{\bold \Omega}}^{11}(q) \equiv \sum_{k} E_k(q) \beta^{\dagger}_k(q) \beta_k(q) \, , \label{onebodypiece}
\end{equation}
is built out of the positive eigenvalues generated through Eq.~\eqref{eq:hfb_equation}. In general, the partitioning defined in Eqs.~\eqref{split1}-\eqref{onebodypiece} is  not {\it canonical}. Indeed, while Eq.~\eqref{canonicalR} is fulfilled for the routhian $R$, it is not for $\Omega$ except for $\lambda_q = 0$, i.e. whenever $| \Phi(q) \rangle$ is the solution of an {\it unconstrained} HFB calculation. 

The BMBPT expansion is formulated using the eigenbasis of $\Omega_{0}(q)$ that is given by
\begin{subequations}
\begin{align}
\Omega_{0}(q)\, |  \Phi(q) \rangle &= {\bold \Omega}^{00}(q) \, |  \Phi(q) \rangle \, , \\
\Omega_{0}(q)\, |  \Phi^{k_1 \ldots}(q) \rangle &= \left[{\bold \Omega}^{00}(q) + E_{k_1}(q)+\ldots\right] |  \Phi^{k_1 \ldots} \rangle  \label{phi} \, .
\end{align}
\end{subequations}

\subsubsection{MBPT}
\label{SRpartitioning2}

Whenever $q_{\text{U(1)}}=0$, $| \Phi(q) \rangle$ is a Slater determinant and BMBPT reduces to MBPT. In this situation, the Bogoliubov field $\bar{\Delta}(q)$ is zero and the Lagrange term associated with the particle number constraint entering the Routhian becomes superfluous and can be omitted. As a result, Eq.~\eqref{eq:hfb_equation} reduces to
\begin{align}
h(q) \, \left(U(q)\right)_k&= e_{k}(q) \, \left(U(q)\right)_k \, , \label{HF1}
\end{align}
i.e. to the constrained HF equation where the one-body HF field reads as
\begin{align}
h^{l}_{l'}(q) &\equiv f^{l}_{l'}[| \Phi(q) \rangle] - \lambda_{q} \frac{\partial {\bold Q}^{00}(q)}{\partial {\rho^*}^{l}_{l'}(q)}  \, . \label{shifted1bodyfield}
\end{align}
Solving Eq.~\eqref{HF1} delivers constrained HF single-particle states $\{a^{\dagger}_k(q)\}$ through the unitary one-body basis transformation $U(q)$ along with the associated HF single-particle energies 
\begin{align}
e_{k}(q) = f^{k}_{k}[| \Phi(q) \rangle]- \lambda_{q} \frac{\partial {\bold Q}^{00}(q)}{\partial {\rho^*}^{k}_{k}(q)}  \, . \label{HFsingpart}
\end{align}
The HF Slater determinant is built by occupying the A lowest HF single-particle states
\begin{align}
| \Phi(q) \rangle &\equiv A^{i_1\cdots i_{\text{A}}}(q) | 0 \rangle \, , \label{product_state_SD}
\end{align}
where a string $A^{p_1\cdots}_{h_1\cdots}(q)$ is defined in terms of constrained HF single-particle creation and annihilation operators.

Because the Lagrange term associated with the particle-number constraint is superfluous, the operator driving the perturbative expansion in MBPT is nothing but the Hamiltonian. Given the above, the unperturbed Hamiltonian deriving from Eq.~\eqref{split2} becomes
\begin{align}
H_{0}(q) &\equiv {\bold H}^{00}(q) + :h(q): \nonumber  \\
&= E^{(0)}(q) + \sum_{k} e_{k}(q) :A^{k}_{k}(q):\ , \label{def_H0_SR1}
\end{align}
where the latter form is given in the eigenbasis of $h(q)$. Equation~\eqref{HFsingpart} makes clear that $h(q)=F_{[| \Phi(q) \rangle]}$ whenever $\lambda_{q}=0$ such that $\{e_{k}(q)\}$ denotes nothing but the eigenvalues of $F_{[| \Phi(q) \rangle]}$ in that particular case.

The $A$-body eigenbasis of $H_0$ is given by
\begin{subequations}
\begin{align}
H_0(q) | \Phi(q) \rangle&=  E^{(0)}(q) | \Phi(q) \rangle  \, , \label{eigenbasis_H0_SR1} \\
H_0(q) | \Phi^{p_1\cdots}_{h_1\cdots}(q) \rangle&=  E^{(0)}_{p_1\cdots h_1\cdots}(q)  | \Phi^{p_1\cdots }_{h_1\cdots}(q) \rangle  \, , \label{eigenbasis_H0_SR2}
\end{align}
\end{subequations}
with 
\begin{subequations}
\begin{align}
E^{(0)}(q) &\equiv {\bold H}^{00}(q) = \langle \Phi(q) | H | \Phi(q) \rangle \, ,\\
E^{(0)}_{p_1\cdots h_1\cdots}(q) &\equiv E^{(0)}(q) + e_{p_1}(q) + \ldots  - e_{h_1}(q) - \ldots \, ,
\end{align}
\end{subequations}
where elementary particle-hole excitations of the unperturbed Slater determinant are defined through
\begin{align}
| \Phi^{p_1\cdots}_{h_1\cdots}(q) \rangle & \equiv A^{p_1\cdots}_{h_1\cdots}(q) | \Phi(q) \rangle \, . \label{elementary_excit_SD}
\end{align}
Whenever applied at the minimum of ${\bold H}^{00}(q)$, the spectrum of $H_0$ is typically non-degenerate with respect to elementary excitations\footnote{The fact that the unperturbed state is non-degenerate is a necessary (but not sufficient) condition for the perturbative series to converge or at least offers mean to be (partially) re-summed. Note that a degeneracy with states carrying different symmetry quantum numbers is not an issue since symmetry blocks are not connected by the perturbation within a symmetry-conserving scheme.}, i.e. it displays a gap-full spectrum
\begin{align}
E^{(0)}_{p_1\cdots h_1\cdots}(q) -E^{(0)}(q)  > 0 \, .
\end{align}
This is schematically illustrated in Fig.~\ref{fig:multiple_rep_HA}.

\subsection{Overlap between Bogoliubov vacua}
\label{overlap}

Given two Bogoliubov vacua \(\ket {\Phi(q,\theta)}\) and \(\ket {\Phi(p)}\), their overlap is a key ingredient to the calculation of the needed many-body matrix elements. To express the result, Bloch-Messiah-Zumino decompositions~\cite{RiSc80} of the Bogoliubov transformations ${\cal W}(p)$ and ${\cal W}(q)$ are invoked, e.g. the matrices defining ${\cal W}(p)$ are expressed as the product of unitary matrices $D(p)$ and $C(p)$ and special block-diagonal matrices $\bar{U}(p)$ and $\bar{V}(p)$ according to
\begin{subequations}
\begin{align}
U(p) &\equiv D(p) \bar{U}(p) C(p) \, , \\
V(p) &= D^*(p) \bar{V}(p) C(p) \, .
\end{align}
\end{subequations}
Further denoting by $v_{k}(p)$ the BCS-like coefficients making up $\bar{V}(p)$, the overlap eventually reads as~\cite{bertsch12}\footnote{There exists an alternative way to compute the overlap between any two Bogoliubov states without any phase ambiguity, see Ref.~\cite{Bally:2017nom}.}
\begin{align}
  \braket{\Phi(p)}{\Phi(q;\theta)} 
  &={}
  (-1)^n
  \frac{\det(C^*(p))\det(C(q))}{\prod_k^n v_{k}(p)v_{k}(q)} 
  \nonumber \\ \times{}&
  \mathrm{pf} \left[
    \begin{pmatrix}
      V(p)^T U(p)     &   V^T(p)  {\bold r}^T(\theta) V^*(q)\\
      -V(q)^\dag  {\bold r}(\theta) V(p)  &   U^\dag(q) V^*(q)
    \end{pmatrix}
    \right] \, , \label{eq:overlap}
\end{align}
where $2n$ denotes the dimension of ${\cal H}_1$ and where the pfaffian of a symplectic matrix has been considered.

\subsection{Transition Bogoliubov transformation}
\label{transitionBgotransfo}

Given the Bogoliubov vacua \(\ket {\Phi(q,\theta)}\) and \(\ket {\Phi(p)}\), the two sets of quasi-particle operators are related via the Bogoliubov transformation
\begin{align}
    \begin{pmatrix}
      \beta(q;\theta)\\ \beta^\dagger(q;\theta)
    \end{pmatrix}
    &=
    \mathcal W^\dag(q;\theta) \mathcal W(p)
    \begin{pmatrix}
      \beta(p)\\ \beta^\dag(p)
    \end{pmatrix}
  \nonumber\\
    &\equiv
    \begin{pmatrix}
      D^\dag(p,q;\theta)  & E^\dag(p,q;\theta)\\
      E^T(p,q;\theta)     & D^T(p,q;\theta)
    \end{pmatrix}
    \begin{pmatrix}
      \beta(p)\\ \beta^\dag(p)
    \end{pmatrix}
  \nonumber\\
    &\equiv \mathcal W^\dag(p,q;\theta)
    \begin{pmatrix}
      \beta(p)\\ \beta^\dag(p)
    \end{pmatrix} \, , \label{transitiontransfoBogo}
\end{align}
where 
\begin{subequations}
 \label{matrixtransitionbogo}
\begin{align}
E(p,q;\theta) &\equiv V^T(q) U(q;\theta) + U^T(q)V(q;\theta)   \, , \\
D(p,q;\theta) &\equiv U^\dagger(q) U(q;\theta) + V^\dagger(q) V(q;\theta) \, .
\end{align}
\end{subequations}
Given that $\mathcal W(p,q;\theta)$ is a unitary Bogoliubov transformation, Eq.~\eqref{unitaryBogo} is also satisfied when substituting $(U(q),V(q))$ for $(D(p,q;\theta),E(p,q;\theta))$.

\subsection{Similarity transformation}
\label{similaritytransfo}

\subsubsection{Thouless transformation}
\label{thouless}

The two Bogoliubov vacua \(\ket {\Phi(q,\theta)}\) and \(\ket {\Phi(p)}\) can be connected via a non-unitary Thouless transformation
\begin{equation}\label{eq:thouless}
  \ket{\Phi(q;\theta)} = \braket{\Phi(p)}{\Phi(q;\theta)} \exp\left[{\bold Z}^{20}(p,q;\theta)\right] \ket{\Phi(p)},
\end{equation}
where matrix elements of the Thouless operator
\begin{equation}
{\bold Z}^{20}(p,q;\theta) \equiv \frac{1}{2}\sum_{k_1k_2} {\bold z}^{k_1k_2}(p,q;\theta) B^{k_1k_2}(p)
\end{equation}
are expressed in terms of the transition Bogoliubov transformation between both vacua (Eqs.~\eqref{transitiontransfoBogo}-\eqref{matrixtransitionbogo}) according to
\begin{equation}\label{eq:thouless_def}
  {\bold z}(p,q;\theta) = E^*(p,q;\theta) D^{*-1}(p,q;\theta) \, .
\end{equation}

\subsubsection{Similarity-transformed operators}
\label{thouless}

Given \(\ket {\Phi(p)}\), \(\ket {\Phi(q,\theta)}\) and an operator $O$, the similarity-transformed operator is introduced as
\begin{equation}
^{Z}O \equiv e^{-{\bold Z}^{20}(p,q;\theta)} O  e^{{\bold Z}^{20}(p,q;\theta)} \, , \label{simtransop}
\end{equation}
which obviously depends on $(p,q;\theta)$ via ${\bold Z}^{20}(p,q;\theta)$. Because the similarity transformation is not unitary, $^{Z}O$ is not hermitian. Such similarity-transformed operators appear repeatedly in the PGCM-PT formalism developed in the present work.

Normal ordering $O$ with respect to \(\ket {\Phi(p)}\) according to Eqs.~\eqref{normalorderedopbogo}-\eqref{defnbodyopbogovacuum}, $^{Z}O$ is obtained by simply replacing the quasi-particle operators \( \{\beta_k^\dagger(p), \beta_k(p)\}\) by the similarity-transformed ones  
\begin{align}
    \begin{pmatrix}
      ^Z\beta(p)\\
      ^Z\beta^{\dag}(p)
    \end{pmatrix}
     &\equiv
    e^{-{\bold Z}^{20}(p,q;\theta)}\begin{pmatrix}
      \beta(p)\\ \beta^\dag(p)
    \end{pmatrix}
    e^{{\bold Z}^{20}(p,q;\theta)}
    \nonumber \\
    &= 
    \begin{pmatrix}
      1 & {\bold z}(p,q;\theta)\\
      0 & 1
    \end{pmatrix}
    \begin{pmatrix}
      \beta(p)\\ \beta^\dag(p)
    \end{pmatrix}
    \nonumber\\
    &\equiv
     {^{Z}\mathcal X}^\dagger(p,q;\theta)
    \begin{pmatrix}
      \beta(p)\\ \beta^\dag(p)
    \end{pmatrix} \, . \label{simtransQPop}
\end{align}
Expressing the result in terms of the initial set \( \{\beta_k^\dagger(p), \beta_k(p)\}\) and applying Wick's theorem allows one to eventually express $^{Z}O$ in normal-ordered form with respect to \(\ket {\Phi(p)}\), i.e. according to Eqs.~\eqref{normalorderedopbogo}-\eqref{stringQP}, where the set of $(p,q;\theta)$-dependent matrix elements are functions of the original set of matrix elements and of the matrix ${\bold z}(p,q;\theta)$. The explicit expressions of these matrix elements are provided in App.~\ref{similaritytransmat} for a two-body operator $O$, i.e. an operator with $r=2$ in Eqs.~\eqref{generalop}-\eqref{stringpart} and/or Eqs.~\eqref{normalorderedopbogo}-\eqref{stringQP}. 

As made clear in App.~\ref{PGCMPT2matel}, one also needs the similarity transformation of a de-excitation operator $B_{l_1 \ldots l_i}(p)$ acting on the corresponding vacuum bra $\bra {\Phi(p)}$, i.e.  
\begin{align}
\bra{\Phi(p)} \, {^{Z}B}_{l_1 \ldots l_i}(p) &
  = \bra{\Phi(p)}  \prod_{n=i}^1  {^{Z}\beta}_{l_n}(p) 
  \\&
  = \bra{\Phi(p)} \prod_{n=i}^1 \left(\beta_{l_n}(p)+\sum_m {\bold z}^{l_il_m}\beta^\dag_{l_m}(p)\right) \, , \nonumber
\end{align}
where the transformation~\eqref{simtransQPop} is used repeatedly and where the dependence of ${\bold z}(p,q;\theta)$ on $(p,q;\theta)$ is omitted for simplicity. This gives for a single de-excitation
\begin{align}
 \bra{\Phi(p)}  \,  {^{Z}B}_{l_1l_2}(p)
  =& \bra{\Phi^{l_1l_2}(p)} \label{eq:ltS-sim-trans} \nonumber \\
  &+ {\bold z}^{l_1l_2} \bra {\Phi(p)} \, , 
\end{align}
and for a double de-excitation 
\begin{align}
 \bra{\Phi(p)}  \,  {^{Z}B}_{l_1l_2l_3l_4}(p)    
 = &\bra{\Phi^{l_1l_2l_3l_4}(p)} \label{eq:ltD-sim-trans} \\
  &+ P(l_1l_2/l_3l_4)  \,  {\bold z}^{l_3l_4}  \, \bra{\Phi^{l_1l_2}(p)}
  \nonumber \\
  &+ P(l_1/l_3l_4)  \, {\bold z}^{l_1l_2}{\bold z}^{l_3l_4}  \,  \bra {\Phi(p)}\, , \nonumber
\end{align}
where the final expressions are obtained by expanding the product of transformed quasi-particle operators, by applying Wick's theorem and by acting on the bra to eliminate many null terms. The definition of the needed permutation operators can be found in App.~\ref{permutop}. 
Interestingly, one observes that the excitation rank is not increased through the similarity transformation in Eqs.~\eqref{eq:ltS-sim-trans}-\eqref{eq:ltD-sim-trans}.

\subsubsection{Rotated/similarity-transformed operators}
\label{thouless}

A rotated/similarity-transformed operator associated to an operator $O$ and given \(\ket {\Phi(p)}\), \(\ket {\Phi(q,\theta)}\) is introduced as
\begin{equation}
^{Z}O(\theta) \equiv e^{-{\bold Z}^{20}(p,q;\theta)} R(\theta) O R^{\dagger}(\theta) e^{{\bold Z}^{20}(p,q;\theta)} \, , \label{simtransop2}
\end{equation}
where the {\it extra} dependence in $\theta$ due to the additional rotation compared to $^{Z}O$ defined in Eq.~\eqref{simtransop} is made apparent in the newly introduced notation $^{Z}O(\theta)$ such that  $^{Z}O(0) = {}^{Z}O$. 

Of course, the particular form used to express the initial operator $O$ does not impact the actual content of $^{Z}O$ or $^{Z}O(\theta)$. In the PGCM-PT formalism of present interest, it happens that $^{Z}O$ and $^{Z}O(\theta)$ arise for operators $O$ that are initially normal  ordered with respect to $| \Phi(p) \rangle$ and $| \Phi(q) \rangle$, respectively, and thus  expressed in terms of quasi-particle operators \( \{\beta_k^\dagger(p), \beta_k(p)\}\) and \( \{\beta_k^\dagger(q), \beta_k(q)\}\), respectively. With this in mind, $^{Z}O(\theta)$ is obtained by simply replacing the quasi-particle operators \( \{\beta_k^\dagger(q), \beta_k(q)\}\) by rotated/similarity-transformed ones 
\begin{align}
    \begin{pmatrix}
      ^Z\beta(q;\theta)\\
      ^Z\beta^{\dag}(q;\theta)
    \end{pmatrix}
        &\equiv
    e^{-{\bold Z}^{20}(p,q;\theta)} R(\theta)
    \begin{pmatrix}
      \beta(q)\\ \beta^\dag(q)
    \end{pmatrix}
    R^{\dagger}(\theta) e^{{\bold Z}^{20}(p,q;\theta)}
    \nonumber\\
        &\equiv
    e^{-{\bold Z}^{20}(p,q;\theta)}
    \begin{pmatrix}
      \beta(q;\theta)\\ \beta^\dag(q;\theta)
    \end{pmatrix}
    e^{{\bold Z}^{20}(p,q;\theta)}
 \, .
       \label{eq:gen_sim_transformed}
\end{align}
 Eventually, the operator $^{Z}O(\theta)$ needs to be re-expressed in terms of the set \( \{\beta_k^\dagger(p), \beta_k(p)\}\), the goal being to express all quantities involved in a many-body matrix element of interest in terms of a single set of quasi-particle operators. To do so, the rotated/similarity-transformed quasi-particle operators are written as  
\begin{align}
    \begin{pmatrix}
      ^Z\beta(q;\theta)\\
      ^Z\beta^{\dag}(q;\theta)
    \end{pmatrix}
        &\equiv
    e^{-{\bold Z}^{20}(p,q;\theta)}
    \begin{pmatrix}
      \beta(q;\theta)\\ \beta^\dag(q;\theta)
    \end{pmatrix}
    e^{{\bold Z}^{20}(p,q;\theta)}
    \nonumber\\
    &=
    \mathcal W^\dag(p,q;\theta)
    {^{Z}\mathcal X}^\dag(p,q;\theta)
    \begin{pmatrix}
      \beta(p)\\ \beta^\dag(p)
    \end{pmatrix}
    \nonumber\\ 
    &\equiv
    {^{Z}\mathcal Y}^\dagger(p,q;\theta)
     \begin{pmatrix}
      \beta(p)\\ \beta^\dag(p)
    \end{pmatrix} \, ,
       \label{eq:gen_sim_transformed}
\end{align}
with 
\begin{align}
{^{Z}\mathcal Y}^\dagger(p,q;\theta)  &=
      \begin{pmatrix}
  D^\dag(p,q;\theta)&
  D^\dag(p,q;\theta) {\bold z}(p,q;\theta)
  +
  E^\dag(p,q;\theta)
\\
  E^T(p,q;\theta)
  & 
  E^T(p,q;\theta) {\bold z}(p,q;\theta) +
  D^T(p,q;\theta) 
  \end{pmatrix} \nonumber \\
  &=        \begin{pmatrix}
  D^\dag(p,q;\theta)&
  0
\\
  E^T(p,q;\theta)
  & 
  D^{*-1}(p,q;\theta) 
  \end{pmatrix}
    \label{eq:gen_sim_transformed2} \, 
\end{align}
where the second line is obtained by inserting Eq.~\eqref{eq:thouless_def} into the first one and utilizing Eqs.~\eqref{unitaryBogoA} and~\eqref{unitaryBogoB}. 

Here, as made clear in App.~\ref{PGCMPT2matel}, one only needs to perform the rotation/similarity-transformation of an excitation operator $B^{k_1 \ldots k_i}(q)$ acting on the vacuum $\ket {\Phi(p)}$ 
\begin{strip}
\vspace{0.8cm}
\begin{align}
{^{Z}B}^{k_1 \ldots k_j}(q;\theta) \, \ket {\Phi(p)} &
  =\prod_{n=1}^j {^{Z}\beta}^{\dagger}_{k_n}(q;\theta) \ket {\Phi(p)}
  \nonumber \\&
  = \prod_{n=1}^j \sum_{k_m}\Big(E_{k_m}^{k_n}\beta_{k_m}(p) 
  + {D^{-1\dag}}^{k_mk_n} \beta_{k_m}^\dag(p)\Big)\ket {\Phi(p)}\, , 
\end{align}
where the transformation~\eqref{eq:gen_sim_transformed}-\eqref{eq:gen_sim_transformed2} has been used repeatedly and where the dependence of $E(p,q;\theta)$ and $D(p,q;\theta)$ on $(p,q;\theta)$ has been omitted for simplicity. This gives for a single excitation
\begin{align}
{^{Z}B}^{k_1k_2}(q;\theta) \, \ket {\Phi(p)} 
  =&   \sum_{j_1j_2} {D^{-1\dag}}^{j_1k_1} {D^{-1\dag}}^{j_2k_2}\ket{\Phi^{j_1j_2}(p)}
    +  \sum_{j_1} E_{j_1}^{k_1}{D^{-1\dag}}^{j_1k_2}\ket{\Phi(p)} \label{rotsimtrans-Sexcit} \, , 
\end{align}
and for a double excitation 
\begin{align}
{^{Z}B}^{k_1k_2k_3k_4}(q;\theta) \, \ket {\Phi(p)} 
  =& \sum_{j_1j_2j_3j_4} {D^{-1\dag}}^{j_1k_1} {D^{-1\dag}}^{j_2k_2} {D^{-1\dag}}^{j_3k_3} {D^{-1\dag}}^{j_4k_4} \ket{\Phi^{j_1j_2j_3j_4}(p)}
   \nonumber\\
   &+
   P(k_1k_2/k_3k_4)\sum_{j_1j_2j_4}
   {D^{-1\dag}}^{j_1k_1} {D^{-1\dag}}^{j_2k_2} E_{j_4}^{k_3} {D^{-1\dag}}^{j_4k_4}
   \ket{\Phi^{j_1j_2}(p)}
   \nonumber\\
   &+
   P(k_1/k_3k_4)\sum_{j_2j_4}
   E_{j_2}^{k_1} {D^{-1\dag}}^{j_2k_2} E_{j_4}^{k_3} {D^{-1\dag}}^{j_4k_4}
   \ket{\Phi(p)} \, , \label{rotsimtrans-Dexcit}
\end{align}
\end{strip}
where the final expressions are obtained by expanding the product of transformed quasi-particle operators, applying Wick theorem and acting on the ket to eliminate many vanishing terms.  Interestingly, one observes that the excitation rank is \emph{not increased} through the rotation and similarity transformation in Eqs.~\eqref{rotsimtrans-Sexcit}-\eqref{rotsimtrans-Dexcit}.

\begin{strip}

\section{Similarity-transformed matrix elements}
\label{similaritytransmat}

Let us consider a two-body operator $O$ (see Eqs.~\eqref{normalorderedopbogo}-\eqref{stringQP} with $r=2$), in normal-ordered form with respect to $| \Phi(p) \rangle$ 
\begin{align}
O \equiv& {\bold O}^{00}(p) \nonumber\\
&+ \Big[{\bold O}^{11}(p) + \{{\bold O}^{20}(p) + {\bold O}^{02}(p)\}\Big]  \nonumber\\
&+ \Big[{\bold O}^{22}(p) + \{{\bold O}^{31}(p) + {\bold O}^{13}(p)\} + \{{\bold O}^{40}(p) + {\bold O}^{04}(p)\}\Big] \nonumber\\
&= {\bold O}^{00}(p) \nonumber\\
&+ \frac{1}{(1!)^2}\sum_{k_1 k_2} {\bold o}^{k_1}_{k_2}(p) B^{k_1}_{k_2}(p) + \frac{1}{2!}\sum_{k_1 k_2} {\bold o}^{k_1 k_2}(p) B^{k_1k_2}(p)  + \frac{1}{2!}\sum_{k_1 k_2} {\bold o}_{k_1 k_2}(p)  B_{k_1k_2}(p)\nonumber\\
&+ \frac{1}{(2!)^{2}} \sum_{k_1 k_2 k_3 k_4} {\bold o}^{k_1 k_2}_{k_3 k_4}(p)
   B^{k_1 k_2}_{k_3 k_4}(p) \nonumber\\
&+ \frac{1}{3!1!}\sum_{k_1 k_2 k_3 k_4} {\bold o}^{k_1k_2k_3}_{k_4}(p)
   B^{k_1k_2k_3}_{k_4}(p) + \frac{1}{1!3!}\sum_{k_1 k_2 k_3 k_4} {\bold o}^{k_1}_{k_2 k_3 k_4}(p) B^{k_1}_{k_2 k_3 k_4}(p)  \nonumber\\
&+  \frac{1}{4!} \sum_{k_1 k_2 k_3 k_4} {\bold o}^{k_1 k_2 k_3 k_4}(p)
   B^{k_1 k_2 k_3 k_4}(p)  +  \frac{1}{4!} \sum_{k_1 k_2 k_3 k_4}  {\bold o}_{k_1 k_2 k_3 k_4}(p)  B_{k_1 k_2 k_3 k_4}(p)   \, . \label{oqpas}
\end{align}
Expressing the similarity-transformed partner $^{Z}O$ (Eq.~\eqref{simtransop}) under the same form, its $(p,q;\theta)$-dependent matrix elements read as
\begin{subequations}
\label{transformrotaME2body1}
  \begin{align}
    ^{Z}{\bold O}^{00} &\equiv ^{Z}{\bold S}^{00}
   \, ,  \\
    {^{Z}{\bold o}}_{k_1k_2} &\equiv {^{Z}s}_{k_1k_2}
   \, ,  \\
    {^{Z}{\bold o}}^{k_1}_{k_2} &\equiv {^{Z}{\bold s}}^{k_1}_{k_2} +
     \sum_{l_1}  {^{Z}{\bold s}}_{l_1k_2} {\bold z}^{l_1k_1}
   \, ,  \\
    {^{Z}{\bold o}}^{k_1   k_2} &\equiv 
    {^{Z}{\bold s}}^{k_1   k_2} 
   +
   P(k_1/k_2)  \sum_{l_1} {^{Z}{\bold s}}^{k_1}_{l_1} {\bold z}^{l_1k_2} 
   -
    \sum_{l_1l_2}  {^{Z}{\bold s}}_{l_1l_2} {\bold z}^{l_1k_1} {\bold z}^{l_2k_2}
    \, , \\
    {^{Z}{\bold o}}_{k_1k_2k_3k_4} &\equiv {^{Z}{\bold s}}_{k_1k_2k_3k_4}
    \, , \\
    {^{Z}{\bold o}}^{k_1}_{k_2k_3k_4} &\equiv 
    {^{Z}{\bold s}}^{k_1}_{k_2k_3k_4}
    + \sum_{l_1} {^{Z}{\bold s}}_{l_1k_2k_3k_4} {\bold z}^{l_1k_1}
   \, ,  \\
    {^{Z}{\bold o}}^{k_1k_2}_{k_3k_4} &\equiv 
    {^{Z}{\bold s}}^{k_1k_2}_{k_3k_4}
    +P(k_1/k_2) \sum_{l_2} {^{Z}{\bold s}}^{k_1}_{l_2k_3k_4} {\bold z}^{l_2k_2}
    - \sum_{l_1l_2}{^{Z}{\bold s}}_{l_1l_2k_3k_4} {\bold z}^{l_1k_1} {\bold z}^{l_2k_2}
   \, ,  \\
    {^{Z}{\bold o}}^{k_1k_2k_3}_{k_4} &\equiv 
    {^{Z}{\bold s}}^{k_1k_2k_3}_{k_4}
    + P(k_3/k_1k_2)\sum_{l_3} {^{Z}{\bold s}}^{k_1k_2}_{l_3k_4} {\bold z}^{l_3k_3}
    \nonumber\\&
    - P(k_1/k_2k_3) \sum_{l_2l_3}{^{Z}{\bold s}}^{k_1}_{l_2l_3k_4} {\bold z}^{l_2k_2}{\bold z}^{l_3k_3}
    - \sum_{l_1l_2l_3}{^{Z}{\bold s}}_{l_1l_2l_3k_4} {\bold z}^{l_1k_1} {\bold z}^{l_2k_2} {\bold z}^{l_3k_3}
    \, , \\
    {^{Z}{\bold o}}^{k_1k_2k_3k_4} &\equiv 
    {^{Z}{\bold s}}^{k_1k_2k_3k_4}
    +P(k_4/k_1k_2k_3) \sum_{l_4}{^{Z}{\bold s}}^{k_1k_2k_3}_{l_4} {\bold z}^{l_4k_4}
    - P(k_1k_2/k_3k_4) \sum_{l_3l_4}{^{Z}{\bold s}}^{k_1k_2}_{l_3l_4} {\bold z}^{l_3k_3} {\bold z}^{l_4k_4}
    \nonumber\\&
    + P(k_1/k_2k_3k_4) \sum_{l_2l_3l_4}{^{Z}{\bold s}}^{k_1}_{l_2l_3l_4} {\bold z}^{l_2k_2}{\bold z}^{l_3k_3}
    {\bold z}^{l_4k_4}
    + \sum_{l_1l_2l_3l_4}{^{Z}{\bold s}}_{l_1l_2l_3l_4} {\bold z}^{l_1k_1} {\bold z}^{l_2k_2} {\bold z}^{l_3k_3}
    {\bold z}^{l_4k_4} \, .
  \end{align}
\end{subequations}
The transformed matrix elements depend on $(p,q;\theta)$ through their dependence on $Z$ and further depend on $p$ through the matrix elements defining the original normal-ordered operator in Eq.~\eqref{oqpas}. All these dependencies have been dropped in Eqs.~\eqref{transformrotaME2body1}-\eqref{transformrotaME2body2} for the sake of readability.

The intermediate matrix elements entering Eqs.~\eqref{transformrotaME2body1}-\eqref{transformrotaME2body2} are defined through
\begin{subequations}
\label{transformrotaME2body2}
  \begin{align}
    ^{Z}{\bold S}^{00} &\equiv
      {\bold O}^{00}
    + \inv2 \sum_{l_1l_2} {\bold o}_{l_1l_2}{\bold z}^{l_1l_2}
     + \inv8 \sum_{l_1l_2l_3l_4} {\bold o}_{l_1l_2l_3l_4} {\bold z}^{l_1l_2}{\bold z}^{l_3l_4}
      \, , \\
    {^{Z}{\bold s}}_{l_1l_2} &\equiv
      {\bold o}_{l_1l_2}
    + \inv2 \sum_{l_3l_4} {\bold o}_{l_1l_2l_3l_4} {\bold z}^{l_3l_4}
     \, ,  \\
    {^{Z}{\bold s}}^{l_1}_{l_2} &\equiv
      {\bold o}^{l_1}_{l_2}
      + \inv2 \sum_{l_3l_4} {\bold o}^{l_1}_{l_2l_3l_4} {\bold z}^{l_3l_4}
      \, , \\
    {^{Z}{\bold s}}^{l_1l_2} &\equiv
      {\bold o}^{l_1l_2}
      + \inv2 \sum_{l_3l_4} {\bold o}^{l_1l_2}_{l_3l_4} {\bold z}^{l_3l_4}
      \, , \\
    {^{Z}{\bold s}}_{l_1l_2l_3l_4} &\equiv
      {\bold o}_{l_1l_2l_3l_4} 
      \, , \\
    {^{Z}{\bold s}}^{l_1}_{l_2l_3l_4} &\equiv
      {\bold o}^{l_1}_{l_2l_3l_4} 
      \, , \\
    {^{Z}{\bold s}}^{l_1l_2}_{l_3l_4} &\equiv
      {\bold o}^{l_1l_2}_{l_3l_4}
     \, ,  \\
    {^{Z}{\bold s}}^{l_1l_2l_3}_{l_4} &\equiv
      {\bold o}^{l_1l_2l_3}_{l_4}
     \, ,  \\
    {^{Z}{\bold s}}^{l_1l_2l_3l_4} &\equiv
      {\bold o}^{l_1l_2l_3l_4}
       \, ,
  \end{align}
\end{subequations}
and incorporate all terms where both indices of a given matrix ${\bold z}$ are contracted with indices of a matrix element of $O$.

\section{PGCM-PT(2) matrix elements}
\label{PGCMPT2matel}

From a technical viewpoint, the building blocks of the PGCM-PT formalism presented in Sec.~\ref{PGCMPT_formalism} are many-body matrix elements of the following kind
	\begin{align}
		O^{\tilde{\sigma}}_{pIqJ} & \equiv \bra{\Phi^I(p)} O P^{\tilde{\sigma}}_{00} \ket{\Phi^J(q)}
		\nonumber                                            
		\\&
		=\frac{d_{\tilde{\sigma}}}{v_{\text{G}}} \sum_\theta
		D_{\text{00}}^{\tilde{\sigma}\ast}(\theta)  \,
		\bra{\Phi(p)} B_{l_1\ldots l_i}(p)  \,O  \,R(\theta)  \,B^{k_1\ldots k_j}(q)| \Phi (q) \rangle 
		\nonumber                                            
		\\&
		=\frac{d_{\tilde{\sigma}}}{v_{\text{G}}} \sum_\theta
		D_{\text{00}}^{\tilde{\sigma}\ast}(\theta)  \,
		\bra{\Phi(p)}  \, {^{Z}B}_{l_1 \ldots l_i}(p) \,\, ^{Z}O \,\, {^{Z}B}^{k_1 \ldots k_j}(q;\theta) \,  | \Phi (p) \rangle  \, \braket{\Phi(p)}{\Phi(q;\theta)}
		\nonumber                                            
		\\&
		\equiv \frac{d_{\tilde{\sigma}}}{v_{\text{G}}} \sum_\theta
		D_{\text{00}}^{\tilde{\sigma}\ast}(\theta)  \,
		O_{pIqJ}(\theta)  \,\braket{\Phi(p)}{\Phi(q;\theta)} \, , \label{genericME}
	\end{align}
where $I$ and $J$ denote arbitrary $i$-tuple and $j$-tuple excitations of the corresponding vacua. Whenever the bra and/or ket is not excited, i.e. whenever the associated excitation operator is the identity, the index is conventionally put to $0$. All quantities appearing in Eq.~\ref{genericME} have been introduced and/or worked out in the previous appendices. 
\end{strip}

While the matrix elements introduced in Eq.~\eqref{genericME} are defined (and could be evaluated) for an operator and excitations of arbitrary ranks, the implementation of PGCM-PT(2) on the basis of a two-body Hamiltonian only requires a subset of them that are now worked out explicitly.

\subsection{Type-1 matrix elements}
\label{MEtwobody}

The first category of matrix elements $O^{\tilde{\sigma}}_{pIqJ}$ is obtained whenever
\begin{enumerate}
\item $O$ is a two-body operator,
\item $I=0,S,D$, where $0/S/D$ stands for no/single/double excitation,
\item $J=0$, i.e. the ket state is fixed to be the vacuum.
\end{enumerate}
Starting thus from a two-body operator $O$ in the form given by Eq.~\eqref{oqpas}, the many-body matrix elements of interest are worked out by exploiting Eqs.~\eqref{eq:ltS-sim-trans}, \eqref{eq:ltD-sim-trans}, \eqref{transformrotaME2body1} and \eqref{transformrotaME2body2} and by applying Wick's theorem with respect to $| \Phi(p) \rangle$ such that
\begin{subequations}
\label{type1ME}
\begin{align}
   O_{p0q0}(\theta)
    &=
    {^Z{\bold O}}^{00} \, ,
    \\
    O_{pSq0}(\theta) 
    &=
    {\bold z}^{k_1k_2} \, {^Z{\bold O}}^{00} + {^Z{\bold o}}^{k_1k_2} \, ,\\
    O_{pDq0}(\theta)
    &=
    P(k_1/k_3k_4){\bold z}^{k_1k_2} {\bold z}^{k_3k_4} {^Z{\bold O}}^{00}
    \\\nonumber
    &+ P(k_1k_2/k_3k_4) {\bold z}^{k_3k_4} {^Z{\bold o}}^{k_1k_2}
    + {^Z{\bold o}}^{k_1k_2k_3k_4} \, ,
\end{align}
\end{subequations}
where the $(p,q;\theta)$ dependencies of the various quantities appearing on the right-hand side have been omitted.

\subsection{Type-2 matrix elements}
\label{MEonebody}

The second category of matrix elements $O^{\tilde{\sigma}}_{pIqJ}$ is obtained whenever
\begin{enumerate}
\item $O$ is a one-body operator,
\item $I=0,S,D$,
\item $J=0,S,D$.
\end{enumerate}
Starting from a one-body operator $O$, i.e. a sub-part of the operator given by Eq.~\eqref{oqpas}, the evaluation of this second category of many-body matrix elements further requires the use of Eqs.~\eqref{rotsimtrans-Sexcit}-\eqref{rotsimtrans-Dexcit} given that excitations of the ket are now in order. 

Vacuum-to-vacuum and $(I=0,J=0)$ and excitation-to-vacuum $(I\neq0,J=0)$ matrix elements can be deduced from Eq.~\eqref{type1ME} and are thus not repeated here. Vacuum-to-excitation $(I=0,J\neq0)$ matrix elements are given by
\begin{strip}
\begin{subequations}
\label{type2ME}
\begin{align}
O_{p0qS}(\theta)
    =&
 {^Z{\bold O}}^{00}  \sum_{j_1} E_{j_1}^{k_1}{D^{-1\dag}}^{j_1k_2}
      \nonumber\\&
+ \sum_{j_1j_2} 
   {^Z{\bold o}}_{j_1j_2} 
   {D^{-1\dag}}^{j_1k_1} {D^{-1\dag}}^{j_2k_2}
    \\
    O_{p0qD}(\theta)
    =&
     P(k_1/k_3k_4)
     {^Z{\bold O}}^{00}
     \sum_{j_2j_4}
   E_{j_2}^{k_1} {D^{-1\dag}}^{j_2k_2} E_{j_4}^{k_3} {D^{-1\dag}}^{j_4k_4}
   \nonumber\\&
   +
    P(k_1k_2/k_3k_4)
    \sum_{j_1j_2j_4}
    {^Z{\bold o}}_{j_1j_2}
   {D^{-1\dag}}^{j_1k_1} {D^{-1\dag}}^{j_2k_2} E_{j_4}^{k_3} {D^{-1\dag}}^{j_4k_4}
   \, .
\end{align}
Excitation-to-excitation  $(I\neq0,J\neq0)$ matrix elements are of course the most involved ones. Single-to-single $(I= S'=\{i_1 i_2\},J= S=\{k_1 k_2\})$ ones read as
\begin{align}
 O_{pS'qS}(\theta) =& 
  P(i_1/i_2)
  {^Z{\bold O}}^{00}         
  {D^{-1\dag}}^{i_1k_1}{D^{-1\dag}}^{i_2k_2}
	\nonumber                                 \\
	&+ 
  P(i_1/i_2)P(k_1/k_2)
	\sum_{j_1}
  {^Z{\bold o}}^{i_1}_{j_1}
  {D^{-1\dag}}^{j_1k_1} {D^{-1\dag}}^{i_2k_2}
	\nonumber                                 \\
	&+ 
  P(k_1/k_2)
	{{\bold z}}^{i_1i_2}
	\sum_{j_1j_2}
  {D^{-1\dag}}^{j_1k_1}
  {^Z{\bold o}}_{j_1j_2}
  {D^{-1\dag}}^{j_2k_2}
	\nonumber                                 \\
	&+
  P(i_1/i_2)
  {^Z{\bold o}}^{i_1i_2}
	\sum_{j_2}
  E_{j_2}^{k_1} {D^{-1\dag}}^{j_2k_2}
	\nonumber                                 \\
	&+
  {^Z{\bold O}}^{00}
  {{\bold z}}^{i_1i_2}
	\sum_{j_2}
  E_{j_2}^{k_1} {D^{-1\dag}}^{j_2k_2} \, .
\end{align}
Double-to-single $(I= D=\{i_1 i_2i_3 i_4\},J= S=\{k_1 k_2\})$ matrix elements read as
\begin{align}
  O_{pDqS}(\theta) &=
  P(i_1i_2 / i_3i_4) P(k_1/k_2) {D^{-1\dag}}^{i_1k_1} {D^{-1\dag}}^{i_2k_2}  {^Z{\bold o}}^{i_3i_4} 
	\nonumber                                             \\
	&+
  P(i_1i_2 / i_3i_4)  P(k_1 / k_2){{\bold z}}^{i_3i_4} {D^{-1\dag}}^{i_1k_1} {D^{-1\dag}}^{i_2k_2} {^Z{\bold O}}^{00}
	\nonumber                                             \\
	&+
	P(i_1/i_2 / i_3i_4)  P( k_1 / k_2 )  {{\bold z}}^{i_3i_4}
    {D^{-1\dag}}^{i_1k_1} \sum_{j_2} {^Z{\bold o}}_{i_2j_2}{D^{-1\dag}}^{j_2k_2} 	
	\nonumber                                             \\
	&+
	P(i_1i_2 / i_3i_4) {{\bold z}}^{i_1i_2}  {^Z{\bold o}}^{i_3i_4}
    \sum_{j_2}E_{j_2}^{k_2} {D^{-1\dag}}^{j_2k_1}	
    \nonumber                                             \\
    &+
	P(i_1i_2 / i_3i_4) {{\bold z}}^{i_1i_2}{{\bold z}}^{i_3i_4} 
    \sum_{j_1j_2} {^Z{\bold o}}_{j_1j_2} {D^{-1\dag}}^{j_1k_1} {D^{-1\dag}}^{j_2k_2}
    \nonumber                                             \\
    &+
	P(i_1/ i_3i_4) {{\bold z}}^{i_1i_2}{{\bold z}}^{i_3i_4} {^Z{\bold O}}^{00}
  \sum_{j_2}E_{j_2}^{k_2} {D^{-1\dag}}^{j_2k_1} \, .
\end{align}
Single-to-double $(I= S=\{i_1 i_2\},J= D=\{k_1 k_2k_3 k_4\})$ matrix elements read as
\begin{align}
  O_{pSqD}(\theta) &=    P(k_1k_2 / k_3k_4) P(i_1 / i_2) {^Z{\bold o}}_{j_3j_4} {D^{-1\dag}}^{j_3k_3}
  {D^{-1\dag}}^{j_4k_4} {D^{-1\dag}}^{i_1k_1} {D^{-1\dag}}^{i_2k_2}
	\nonumber                                             \\
	&+
  P(k_1k_2 / k_3k_4)P(i_1 / i_2) 
    {^Z{\bold O}}^{00} {D^{-1\dag}}^{i_1k_1} {D^{-1\dag}}^{i_2k_2}\sum_{j_3} E_{j_3}^{k_4}{D^{-1\dag}}^{j_3k_3}
	\nonumber                                             \\
	&+
  P(k_1 / k_2 / k_3k_4) P(i_1 / i_2) {D^{-1\dag}}^{i_1k_1} \left(\sum_{j_3} E_{j_3}^{k_3} {D^{-1\dag}}^{j_3k_4}\right)
  \left(\sum_{j_2} {D^{-1\dag}}^{j_2k_2} {^Z{\bold o}}^{i_2}_{j_2} \right) 
	\nonumber                                             \\
	&+
  P(k_1k_2 / k_3k_4) {{\bold z}}^{i_2i_1} \left(\sum_{j_3} E_{j_3}^{k_4} {D^{-1\dag}}^{j_3k_3}\right)
  \left(\sum_{j_1j_2} {^Z{\bold o}}_{j_1j_2} {D^{-1\dag}}^{j_1k_1} {D^{-1\dag}}^{j_2k_2} \right) 
	\nonumber                                             \\
	&+
P(k_1/ k_3k_4) {^Z{\bold o}}^{i_1i_2}
  \left(\sum_{j_1} E_{j_1}^{k_1} {D^{-1\dag}}^{j_1k_2}\right)
  \left(\sum_{j_3} E_{j_3}^{k_4} {D^{-1\dag}}^{j_3k_3}\right)
	\nonumber                                             \\
	&+
  P(k_1/ k_3k_4) {^Z{\bold O}}^{00} {{\bold z}}^{i_2i_1}
	 \left(\sum_{j_3}
  E_{j_3}^{k_4} {D^{-1\dag}}^{j_3k_3}\right)
  \left(\sum_{j_1}
  E_{j_1}^{k_2} {D^{-1\dag}}^{j_1k_1}\right)
	\, .
\end{align}
Double-to-double $(I= D=\{i_1 i_2i_3 i_4\},J= D'=\{k_1 k_2k_3 k_4\})$ matrix elements read as
\begin{align}
  O_{pDqD'}(\theta) &= P(i_4/i_3/i_2/i_1)
  {^Z{\bold O}}^{00}
  {D^{-1\dag}}_{k_4i_4} {D^{-1\dag}}^{i_3k_3}
  {D^{-1\dag}}^{i_2k_2} {D^{-1\dag}}^{i_1k_1}
	\nonumber                                             \\
	&+
  P(k_4/k_3k_2k_1)
  P(i_4/i_3/i_2/i_1)
  \left(
  \sum_{j_4}
  {^Z{\bold o}}^{i_4}_{j_4}
  {D^{-1\dag}}^{j_4k_4}
  \right)
  {D^{-1\dag}}^{i_3k_3} {D^{-1\dag}}^{i_2k_2} {D^{-1\dag}}^{i_1k_1}
	\nonumber                                             \\
	&+ 
	P(i_1i_2/i_3i_4)
	{{\bold z}}^{i_1i_2}
  P(k_1k_2/k_3/k_4)
  {D^{-1\dag}}^{i_4k_4} {D^{-1\dag}}^{i_3k_3}
	\sum_{j_2j_1}
  {D^{-1\dag}}^{j_1k_1} {^Z{\bold o}}_{j_1j_2}
  {D^{-1\dag}}^{j_2k_2}
	\nonumber                                             \\
	&+
	P(k_1k_2/k_3k_4)
	P(i_1i_2/i_3/i_4)
  {^Z{\bold o}}^{i_1i_2}
  {D^{-1\dag}}^{i_4k_4}
  {D^{-1\dag}}^{i_3k_3}
  \left(
	\sum_{j_1}
  E_{j_1}^{k_2} {D^{-1\dag}}^{j_1k_1}
  \right)
	\nonumber                                             \\
	&+
	P(i_1i_2/i_3/i_4)
	P(k_1k_2/k_3k_4)
	{{\bold z}}^{i_1i_2}
  {^Z{\bold O}}^{00}         
  {D^{-1\dag}}^{i_4k_4} {D^{-1\dag}}^{i_3k_3} \left(\sum_{j_1}
  E_{j_1}^{k_2} {D^{-1\dag}}^{j_1k_1}
  \right)
	\nonumber                                             \\
	&+
	P(i_1i_2/i_3/i_4)
	P(k_1k_2/k_3/k_4)
	{{\bold z}}^{i_1i_2}
  \left(\sum_{j_1}
  E_{j_1}^{k_2} {D^{-1\dag}}^{j_1k_1}
  \right)
 \left(	\sum_{j_4}
  {^Z{\bold o}}^{i_4}_{j_4}
  {D^{-1\dag}}^{j_4k_4} {D^{-1\dag}}^{i_3k_3}  \right)
	\nonumber                                             \\
	&+
	P(i_1/i_3i_4)
	P(k_1k_2/k_3k_4)
    {{\bold z}}^{i_1i_2}
	{{\bold z}}^{i_3i_4}
  \left(\sum_{j_1}
  E_{j_1}^{k_2} {D^{-1\dag}}^{j_1k_1}
  \right)
	\left(\sum_{j_4j_3}
  {D^{-1\dag}}^{j_3k_3}
  {^Z{\bold o}}_{j_3j_4}
  {D^{-1\dag}}^{j_4k_4}  \right)
	\nonumber                                             \\
	&+
	P(i_1i_2/i_3i_4)
	P(k_1/k_3k_4)
	{{\bold z}}^{i_1i_2}{^Z{\bold o}}^{i_3i_4}
  \left(
	\sum_{j_3}
  E_{j_3}^{k_4} {D^{-1\dag}}^{j_3k_3}
  \right)\left(
	\sum_{j_1}
  E_{j_1}^{k_2} {D^{-1\dag}}^{j_1k_1}
  \right)
	\nonumber                                             \\
	&+
	P(i_1/i_3i_4)
	P(k_1/k_3k_4)
	{{\bold z}}^{i_1i_2}
	{{\bold z}}^{i_3i_4}
	{^Z{\bold O}}^{00}
  \left(
	\sum_{j_3}
  E_{j_3}^{k_4} {D^{-1\dag}}^{j_3k_3}
  \right)\left(
	\sum_{j_1}
  E_{j_1}^{k_2} {D^{-1\dag}}^{j_1k_1}
  \right)
   \, .
\end{align}
\end{subequations}
\end{strip}

\subsection{Type-3 matrix elements}
\label{MEzerobody}

The third category of matrix elements is obtained from $O^{\tilde{\sigma}}_{pIqJ}$ whenever
\begin{enumerate}
\item $O$ is a zero-body operator, i.e. the identity operator multiplied by the number $O^{00}$,
\item $I=0,S,D$,
\item $J=0,S,D$.
\end{enumerate}
All these matrix elements can be deduced from the previous cases by solely keeping the terms proportional to ${^Z{\bold O}}^{00} = O^{00}$ in the appropriate expressions.

\end{appendices}

\bibliography{bibliography.bib}

\end{document}